\documentclass[%
 reprint,
 amsmath,amssymb,
 aps,
prl,twocolumn,longbibliography,citeautoscript
]{revtex4-2}

\usepackage{braket}
\usepackage{caption}
\usepackage{quantikz}
\usepackage{graphicx}
\usepackage{dcolumn}
\usepackage{bm}
\usepackage{xcolor}
\definecolor{apsblue}{rgb}{0,0,0.8}
\usepackage[colorlinks=true,linkcolor=apsblue, citecolor=apsblue, urlcolor=apsblue]{hyperref}
\usepackage{comment}
\usepackage{booktabs}

\newcommand{\fullfigref}[2]{\hyperref[#1]{\ref*{#1}(#2)}}

\usepackage{tikz}

\newcommand{\tikzcircle}[2][red,fill=red]{\tikz[baseline=-0.6ex]\draw[#1,radius=#2] (0,0) circle ;}

\begin{document}
\nocite{apsrev42Control}

\title{Quantum Correlations and Entanglement in Generalized Dicke-Ising Models}

\author{Santiago F. Caballero-Benitez$^{1}$}%
 \email{scaballero@fisica.unam.mx}
\affiliation{%
$^{1}$Instituto de Física, LSCSC-LANMAC, Universidad Nacional Autónoma de México, Ciudad de México 04510, Mexico
}%

\setcitestyle{super}
\makeatletter
\renewcommand\@biblabel[1]{\textsuperscript{#1.}}
\makeatother
\begin{abstract}
Quantum systems inside high-Q cavities offer an excellent testbed for the control of emergent symmetries induced by light and their  interplay with quantum matter.
Recently several developments in cavity experiments with neutral atoms and other quantum objects such as ions motivate the study of their quantum correlated properties and their entanglement to tailor and control 
the behavior of the system. Using the enhanced coupling between light and interacting matter we explore the properties of emergent superradiant modes using our newly developed Light-Matter DMRG algorithm with strongly interacting spin chains. We explore a experimentally viable generalization of the transverse Ising chain coupled to the cavity light where it is possible to induce multimode structures tailored by the light pumped into the system. We find a plethora of scenarios can be explored with clear and accesible measurable signatures. This allows to study the physics of emergent orders and strong quantum correlations with quantum spins where the local and long range coupling can be efficiently simulated. We find that quantum spin nematic states with long range order and magnon pairs emerge as the transitions to superradiant phases take place. 
Notably, we show the cavity field allows the optimization of entanglement between spins for different light induced modes which can be used for quantum state engineering of quantum correlated states. Our methods can be used to model other hybrid quantum systems efficiently.

\end{abstract}

\maketitle

Quantum scatterers inside high-Q cavities or confined in waveguides offer a unique opportunity to study how strong light-matter coupling can influence the emergence of macroscopic ordering and how quantum correlations determine the properties of the system. Recent advances in cavity systems with ultracold atoms~\cite{doi:10.1080/00018732.2021.1969727}, motivate to explore in other systems such as effective spins, how the properties of the system can be controlled. Using the geometry of light it is possible to induce modes that collectively scatter light in groups across the system~\cite{PhysRevLett.114.113604,PhysRevA.93.023632,PhysRevLett.115.243604}. These groups of quantum objects will have collective emission that can be exploited for quantum information purposes. Moreover, as light matter interaction becomes strong inside the cavity,  superradiance (SR) emerges~\cite{Baumann2010,PhysRevResearch.3.L012024} triggering a quantum phase transition~\cite{Sachdev2011} while modifying the collective properties of the scatterers due to cavity back action. Thus, macroscopic order emerges controlled by how we pump light into the system by design. In an array of these emitters, we can have that different macroscopic orders compete similar to atomic systems where several quantum correlated states have been observed, such as insulators and supersolids~\cite{Landig2016,Hemmerich,Leonard2017}, while magnetic ordering is also possible~\cite{PhysRevLett.121.163601,PhysRevLett.128.080601,Brahyam}. In effective spin systems inside cavities such as neutral atoms~\cite{Rey,PhysRevResearch.3.013178}, Rydberg atoms~\cite{CavityRydberg,Rydberg1,SciPostPhys.1.1.004}, ions in wave guides~\cite{ions1,ions2} using the motional degrees of freedom and in atoms near waveguides~\cite{PhysRevResearch.6.023213,Solano2017}, the situation is similar structurally which motivates our work. 
\begin{figure}[ht!]
    \centering
 \includegraphics[width=0.49\textwidth]{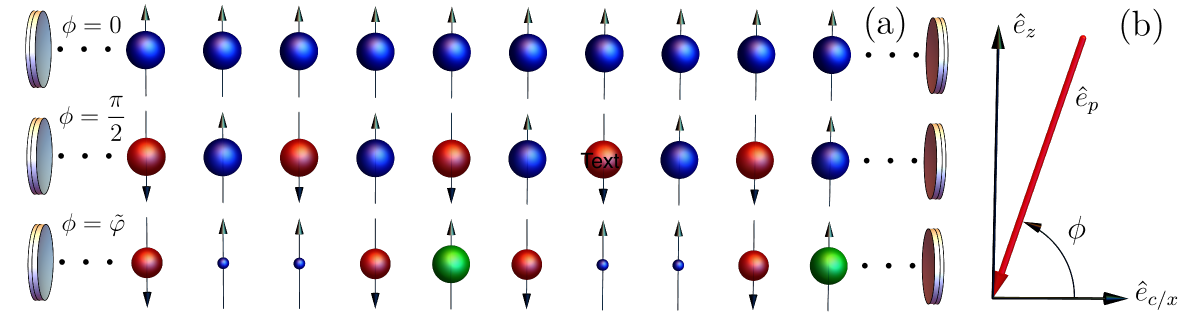}
    \caption{
     \textbf{Light mode structure induced to the spin chains in the cavity}
     \textbf{(a)}-\textbf{(b)}, Experimental scheme for the light.  Panel \textbf{(a)}: Spins along the chain, sphere size is proportional to the light mode coupling strength at each site. The diffraction maxima configuration ($\phi=0$), a ferromagnetic mode is induced [\tikzcircle[fill=blue]{3pt}\tikzcircle[fill=blue]{3pt}]. The diffraction minima configuration ($\phi=\pi/2$), generates a staggered spin field  [\tikzcircle[fill=red]{3pt}\tikzcircle[fill=blue]{3pt}]. 
     The golden ratio (RBBRG) configuration ($\phi=\arccos(1/5)=\tilde{\varphi}$) inducing a three mode staggered spin structure (red, blue, green) every 5 sites  [\tikzcircle[fill=red]{3pt}\tikzcircle[fill=blue]{3pt}\tikzcircle[fill=blue]{3pt}\tikzcircle[fill=red]{3pt}\tikzcircle[fill=green]{3pt}]. The arrows point the direction of the spin in the quantization axis ``$x$". Panel \textbf{(b)}: The light setup,  the angle  $\phi$ between the cavity axis $\hat{e}_c$ and the light pumped into the system with direction $\hat{e}_p$ ($-\hat{e}_p$) for a standing wave cavity and retro-reflected pump beams in the pump axis from the side (only one is shown, the other one is parallel).  The light polarization unit vector is $\hat{e}_z$ and the spins are along the spatial axis $\hat{e}_x$. 
}    
\label{fig:Fig1}
\end{figure}
With this in mind, we study a prototypical setup to explore the collective properties of interacting spins with its interplay with cavity light in a one dimensional setup. The interaction mediated by the cavity effective couples all the spins in the system with a global structured interaction that competes with the underlaying structure that the spins have. The spins in their most simple form can interact with each other as in the Ising model, providing the system with effectively an anharmonic deformation to the energy structure of the system. The fact that now the system has many body interactions  in two fronts via the cavity light and the direct spin interactions will trigger competition of orders akin to the situation in ultracold atomic systems. As a consequence of this, quantum phase transitions emerge and the SR phases can have non-trivial properties. Further, we present a rich landscape of quantum many body phases for the spin ordering with useful properties in the structures seeded by light for quantum information purposes as the entanglement can be controlled by design.  For the simulations of the system, we employ the Light-DMRG algorithm~\cite{gvm2-b46t}, see SM~\cite{SM}. Recently, similar methods have been applied with the single mode case~\cite{DIp} and other methods have been studied~\cite{lcvj-ksct,PhysRevResearch.2.023131}.

\begin{table}[]
    \begin{tabular}{ c c c c c c c c c}
\toprule
\toprule
        Identifier  & Phase & $|\braket{\hat a}|$ & $\mathcal{M}^z_0$ & $\mathcal{M}^x_0$ & $\mathcal{M}^z_\pi$ & $\mathcal{M}^x_\pi$ & $
      \tilde{\mathcal{Q}}^B$ & $\tilde{\mathcal{P}}$   \\
    \bottomrule
        I      & N-FM$_z$                                               & 0              &  $\neq 0$  &  0            &  0             &     0          & 0              & 0\\
        
        II     & N-AFM$_z$                                             & 0              &       0         & 0             &  $\neq 0$ & 0             & 0              & 0\\
        
        III    & SR-FM$_x$                                        &  $\neq 0$ &     0           &  $\neq 0$  &  0               & 0             & $\approx 0$               & $\neq 0$\\    
        
        IV   & SR-AFM$_x$                                     &   $\neq 0$ &    $\neq 0$            & 0              &  0              & $\neq 0$ & $\approx 0$             &  $\neq 0$\\
          
        V  & SR-FM$_x$-$\mathcal{Q}^B$           &  $\neq 0$  &0                 & $\neq 0$ &  0              & 0            &  $\neq 0$ & $\neq 0$ \\    
        
        VI & SR-AFM$_x$-$\mathcal{Q}^B$         &  $\neq 0$ &    0             &  0             &  0             & $\neq 0$ & $\neq 0$ & $\neq 0$ \\
        
        VII   & SR-FM$_\varphi$-$\mathcal{Q}^B$    &  $\neq 0$ &     $\neq 0$           & $\neq 0$   & $\approx 0$ & $\approx 0$ & $\neq 0$ &  $\neq 0$\\

     \hline
     \hline
    \end{tabular}
    \caption{{\textbf{Quantum phases of the spins and their order parameters.}} The order parameters are the amplitude $|\braket{\hat a}|$ of the light field, the magnetization $\mathcal{M}_0^\nu$, staggered magnetization $\mathcal{M}_\pi^\nu$, the mean  bond nematic order parameter $\tilde{\mathcal{Q}}^B$, the magnon pair order parameter $\tilde{\mathcal{P}}$ and the dominant spin axis $\nu=x/z$. In the normal phases (N), $|\langle\alpha\rangle|=0$, there is no collective photon emission.
    The resulting phases are combinations of normal N or superradiant order R, the dominant spin axis induced by the light and how it is transferred to the spins generating FM/ AFM in ``$x$" or Multimode FM order with golden ratio coupling $\varphi$, the bond quadrupolar order in the spins and the emergence of magnon pairs with long range order.}
    \label{tab:phases}
\end{table}

\begin{figure}
    \centering
    \includegraphics[width=0.47\textwidth]{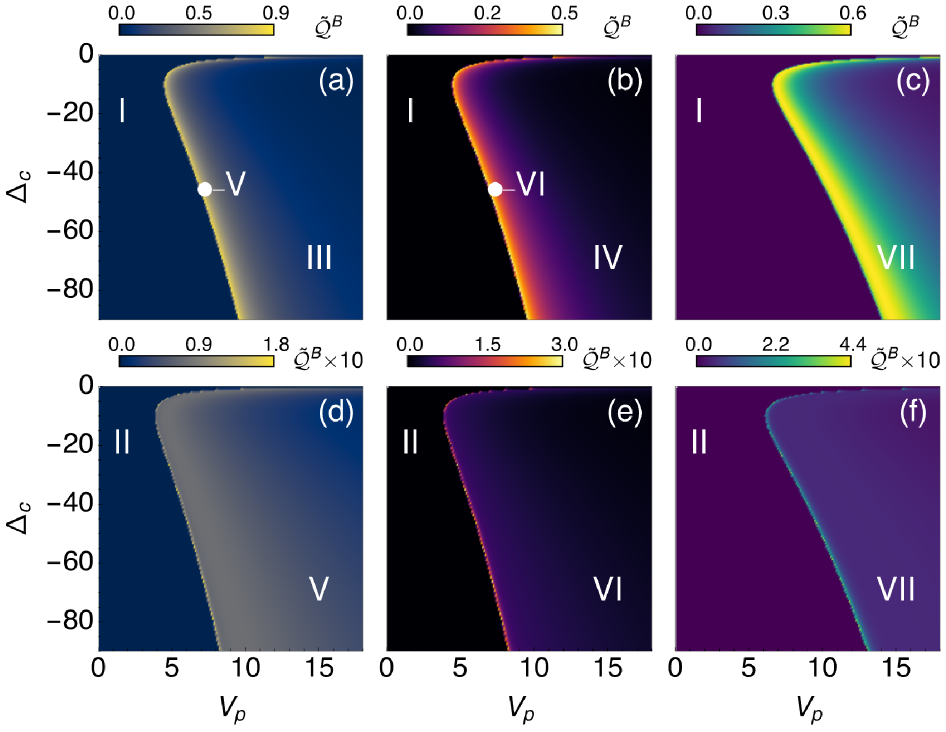}
    \caption{
     \textbf{Phase diagrams for different light induced mode structures and the behavior of the mean bond-nematic order parameter ($\tilde{\mathcal{Q}}^B$)}.
     When the system becomes superradiant and the quantization axis of the spins changes there is the formation of  bond nematic quantum states. These can persists deeper into the superradiant state depending on the short-range spin interaction strength $J$ and the light induced configuration.
     Panels (\textbf{a},\textbf{b},\textbf{c}): correspond to $J<0$, normal FM state (I)  and superradiant configurations (III-VII).
     Panels (\textbf{d},\textbf{e},\textbf{f}):  correspond to $J>0$  normal AFM state (II) and superradiant configurations (V-VII). The light induced mode structures  are: (\textbf{a},\textbf{d}) for $\phi=0$ diffraction maxima [\tikzcircle[fill=blue]{3pt}\tikzcircle[fill=blue]{3pt}], (\textbf{b},\textbf{e}) for $\phi=\pi/2$ diffraction minima [\tikzcircle[fill=red]{3pt}\tikzcircle[fill=blue]{3pt}] and (\textbf{c},\textbf{f}) for $\phi=\arccos(1/5)$, the golden ratio configuration [\tikzcircle[fill=red]{3pt}\tikzcircle[fill=blue]{3pt}\tikzcircle[fill=blue]{3pt}\tikzcircle[fill=red]{3pt}\tikzcircle[fill=green]{3pt}], as shown in  Fig.\ref{fig:Fig1} (a). The order parameters of different phases are in Table \ref{tab:phases}. 
     Parameters in the simulations are: $N=400$ spins, $\omega_0/|J|=0.1$, $\kappa/|J|=10$ with $\Delta_c$ and $V_p$ in units of $|J|$. 
    }
    \label{fig:Fig2}
\end{figure}

{\it The model.} The many body Hamiltonian of the system rotating at frequency $\omega_p$, following analog atomic systems methods \cite{Maschler2008,NewJPhys.17.123023} is:
\begin{eqnarray}
\mathcal{H}&=&-\hbar\Delta_c\hat a^\dagger\hat a+\hbar\omega_0 \sum_n\hat{S}^z_n
+\hbar J\sum_{n}\hat{S}^z_n\hat{S}^z_{n+1}
\nonumber\\
&+&\frac{\hbar V_p}{\sqrt{N}}\left(\hat{a}^\dagger+\hat{a}\right)\sum_nJ^{pc}_n\hat{S}^x_n
\label{model}
\end{eqnarray}
where  the first term is the energy of the cavity photons, $\Delta_c = \omega_p-\omega_c$ is the cavity detuning, $\hbar$ the reduced Planck constant, $\hat a^{(\dagger)}$ is the cavity mode photonic annihilation (creation) operator. The energy of each spin $1/2$ is $\hbar\omega_0$, the interaction between spins in the absence of cavity light is given by the Ising Hamiltonian with $\hbar J$ the energy of the exchange of spins.  The interaction between the photons and the spins is given in the typical Dicke form~\cite{PhysRevA.100.013816,PhysRevLett.121.163601,doi:10.1080/00018732.2021.1969727} with $\hbar V_p$ the effective pump potential, the overlap amplitudes of the spins with the pump and cavity light are given by $J^{pc}_n=\cos(\pi n)\cos(\pi n\cos\phi)$ at each localized spin with site $n$ and $N$ spins; the angle between the pump wave wave vector $\mathbf{k}_p=|\mathbf{k}_p|\hat{e}_p$ and the cavity wavevector $\mathbf{k}_c=|\mathbf{k}_c|\hat{e}_c$ is $\phi$, the spins are along $\hat{e}_x$ as shown in Fig.\ref{fig:Fig1},  see SM~\cite{SM}. Similar to atomic systems~\cite{PhysRevLett.114.113604,PhysRevA.93.023632,PhysRevLett.115.243604}, we consider three different configurations  depending on the light induced emergent coupling of the spins due to the cavity field, for $\phi=0$ the coupling is homogenous across the lattice inducing a single mode, with $\phi=\pi/2$ we induce a two mode staggered ordering with couplings $J^{pc}_n=(-1)^n$ and with $\phi=\arccos(1/5)\approx0.436\pi$, the golden ratio mode, we induce a pattern (RBBRG) for the couplings $J^{pc}_n=\{-\varphi/2,(\varphi-1)/2,(\varphi-1)/2,-\varphi/2,1\}$ every 5 sites with $\varphi\approx 1.61803$ the golden ratio. The particular choice of the golden ration mode is motivated by the fact that this configuration is the minimal case where couplings have just three different non zero values and the structure allows for these three modes to scatter light with the same phase for each group of sites across the whole lattice. Indeed, other choices with this property with more modes at different angles can be found as we will see below. Additional choices with sets with groups of sites without coupling to the cavity light have been explored in~\cite{PhysRevLett.114.113604,PhysRevA.93.023632,PhysRevLett.115.243604}. To achieve a conmensurable pattern of these modes reliably, the precision in the control of the angle scales as $|\delta\phi|\gtrsim 10^{-2}/N$, for an accuracy of $S(\delta\phi)\sim10^{-2}$, see the SM~\cite{SM} for details.

{\it The order parameters.} In spin systems one body order parameters that characterize the magnetic ordering in the system are the magnetizations $\mathcal{M}^{\nu}_\theta=\sum_n\cos(\theta n)\langle\hat{S}^\nu_n\rangle/N$, with $\nu\in\{x,y,z\}$ and $\theta=0$ for the magnetization (FM ordering) or $\theta=\pi$ the staggered magnetization (AFM ordering) in the spin axis $\nu$.
Considering genuine many body spin quantum correlations, for spin $1/2$ systems the traceless bond nematic tensor of rank 2 is defined as:
\begin{equation}
\mathcal{Q}^{\nu\eta}_{nm}=\langle\hat S^\nu_n\hat S^\eta_m+\hat S^\eta_n\hat S^\nu_m\rangle(1-\delta_{\nu\eta})(1-\delta_{nm})/2
\end{equation}
for each pair of sites $nm$, with $\nu,\eta \in \{x,y,z\}$. This tensor measures the quadrupolar deformations in the system between different sites as the on-site signal is trivial for spin $1/2$ systems\cite{PhysRevLett.96.027213,PhysRevLett.130.116701}. The bond nematic order parameter for each pair of sites $nm$ is the largest eigenvalue of the tensor such that $\mathcal{Q}^B_{nm}=\max(\mathrm{Sp}(\mathcal{Q}_{nm}))$. It is useful to define the average over all the pairs of sites in the lattice as the global order parameter of how nematic is the state. Thus, we define the mean bond nematic order parameter as $\tilde{\mathcal{Q}}^B=\sum_{nm}\mathcal{Q}^B_{nm}/N^2$, this is shown in Fig.\ref{fig:Fig2} for the different light induced modes in Fig.\ref{fig:Fig1} and different Ising couplings (FM/AFM). Due to the ``$x$-$z$" symmetry of $\mathcal{H}$, the nematic order  is uniaxial with $\mathcal{Q}^B_{nm}=|\mathcal{Q}^{xz}_{nm}|\neq 0$ whenever it emerges. Clearly, this is an effect that cannot be obtained by mean-field calculations and it clarifies the underlying mechanism that triggers the quantum phase transition due to strong quantum correlations. In addition to the nematic character of the states, it is well known that magnon pair states can be supported with nematic ordering~\cite{PhysRevB.44.4693,PhysRevLett.96.027213}, the magnon pairs at different sites are characterized by pair amplitude $\mathcal{P}_{nm}=\langle \hat{S}^-_n\hat{S}^-_m\rangle=\langle\hat{S}^x_n\hat{S}^x_m-\hat S^y_n\hat S^y_m-i\hat{S}^x_n\hat{S}^y_m-i\hat{S}^y_n\hat{S}^x_m\rangle$ and the global magnon pair order parameter can be defined as $\tilde{\mathcal{P}}=\sum_{nm}|\mathcal{P}_{nm}|/N^2$.

{\it Quantum phases.} We find a rich set of scenarios that can be tailored depending on the geometry of the light and the short range Ising interaction present with either FM$_z$ or AFM$_z$ character. The underlaying Ising interaction determines how the correlations induced by the cavity light form. The SR phases in the system ($|\langle\alpha\rangle|\neq0$) can be FM$_x$/AFM$_x$  [Fig.\ref{fig:Fig2} (a,d)/(b,e)] with  weak nematic character for large $V_p$. In contrast,  fully nematic order occurs with the FM$_\varphi$ pattern induced to the system  [Fig.\ref{fig:Fig2} (c,f)].  All the phases present are shown in  Fig.\ref{fig:Fig2} and the Table \ref{tab:phases} has the definitions and behavior of the order parameters for each quantum phase.  Further details of all relevant order parameters can be found in the SM~\cite{SM}. We find that as the cavity light emerges, the spins twist their quantization axis leading to the formation of a uniaxial  quantum spin nematic~\cite{PhysRevB.109.L220407} as the clear signature of the quantum phase transition. Moreover, the presence of these states can be stabilized beyond the transition region with the $\varphi$ light induced spin modes in both underlying short range FM$_z$ and AFM$_z$ couplings Fig. \ref{fig:Fig2} (c,f). The difference between SR phases with and without nematic character originates from the emergence of quantum correlations in the nematic phase. When $\tilde{\mathcal{Q}}^B=0$, we have only dipolar order (canted magnet) as $\langle \hat S^x_n\rangle\neq0$, the system has minimal quantum correlations. From our simulations, we find that the FM$_z$ Ising interactions are better for the stabilization of bond nematic order. Intuitively, this can be understood as the rigidity of the underlying FM$_z$ state is larger with respect to the twist applied to the spin quantization axis by the cavity light with respect to the AFM$_z$ state that is already minimizing the total spin in the system in the normal state. The quantum phase transitions are strongly dependent on the light and cavity parameters, modifying the cavity detuning and the pump strength it is possible to explore how the quantum phase transitions take place in a single setup. Surprisingly, we find that all the SR states support magnon pairs with quasi-long range order, even with a minimal display ($\tilde{\mathcal{Q}}^B\approx 0\neq0$) of nematic order. Further, we perform finite size analysis across the system of the quantum covariance of spin projections, which for our system are $\delta\mathcal{Q}_{nm}^B=\langle \hat S_n^x\hat S_m^z\rangle-\langle \hat S^x_n\rangle\langle\hat S_m^z\rangle$, we find that indeed our system has true long range bond nematic order ($\delta\mathcal{Q}_{nm}^B\neq0$) coexisting with dipolar order as $\langle\hat S_n^x\rangle\neq 0$ simultanously, see the SM~\cite{SM} for details of the analysis. In contrast, performing the same analysis for the magnon pair amplitude with $\delta\mathcal{P}_{nm}=\langle \hat S_n^x\hat S_m^x\rangle-\langle \hat S^x_n\rangle\langle\hat S_m^x\rangle$, reveals that the transverse spin fluctuations exhibit quasi-long range order. Measuring the cavity light amplitude $|\langle\hat a\rangle|$, the magnetizations  and the spin correlations, similar to~\cite{PhysRevLett.121.163601,PhysRevLett.132.093402}, it is experimentally viable to reconstruct the phase diagrams of Fig. \ref{fig:Fig2}. In similar setups, these are routinely produced experimentally~\cite{doi:10.1080/00018732.2021.1969727}. For additional details on all the observables see the SM~\cite{SM}.

\begin{figure}[t]
    \centering
    \includegraphics[width=0.47\textwidth]{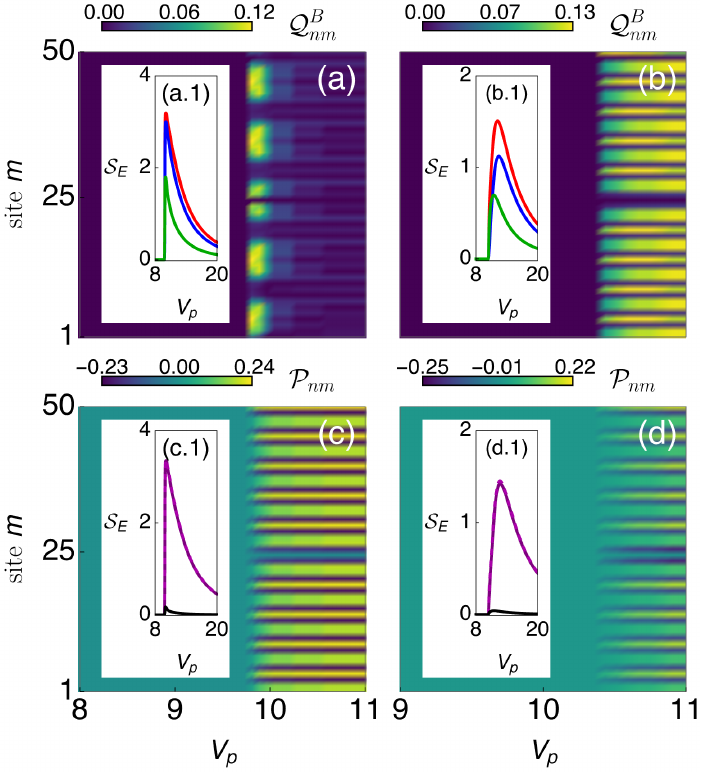}
    \caption{
     \textbf{Behavior with the golden ratio mode $\varphi$ induced by light. Main Panels: Local bond nematic order parameter $\mathcal{Q}^B_{nm}$, local magnon pair amplitude $\mathcal{P}_{nm}$. Inset Panels: Entanglement entropies $\mathcal{S}_E$ for different mode structures in the system.}
     Main panels (\textbf{a},\textbf{b}) show the amplitude variation of $\mathcal{Q}^B_{nm}$ near the transition to the SR state and how long range structured correlations emerge across the whole lattice. Main panels (\textbf{c},\textbf{d}) show the amplitude variation of the magnon pairs with long range coherent oscillations in the SR. The Ising coupling is $J>0$ for (a, c) and $J<0$ for (b, d). Inset panels  (\textbf{a.1},\textbf{b.1}) show the entanglement entropy of the sites with R modes (red) [\tikzcircle[fill=red]{3pt} - - \tikzcircle[fill=red]{3pt} -], the B modes (blue) 
     [ - \tikzcircle[fill=blue]{3pt} \tikzcircle[fill=blue]{3pt}  - -], and the G mode (green)  [- - - - \tikzcircle[fill=green]{3pt}], across the whole lattice, following Fig.\ref{fig:Fig1}.  Inset panels  (\textbf{c.1},\textbf{d.1}) show the entanglement entropy of the sites AFM$_x$ singlet sites ($1+k$ and $2+k$) [solid purple, \tikzcircle[fill=red]{3pt} \tikzcircle[fill=blue]{3pt} - - -] and ($3+k$ and $4+k$) [dashed  purple, - - \tikzcircle[fill=blue]{3pt} \tikzcircle[fill=red]{3pt}  -] across the whole lattice  with $k=5\times (n-1)$, $n=1,\dots,N/5$; the entanglement entropy of half the chain is shown in black. The parameters are: $n=25$ the middle of the chain for $\mathcal{Q}^B_{nm}$ and $\mathcal{P}_{nm}$; $N=50$ sites, $\omega_0/|J|=0.1$, $\kappa/|J|=10$, $\Delta_c/|J|=-50$ and $V_p$ in units of $|J|$ for all plots. 
     }
    \label{fig:Fig3}
\end{figure}

{\it Local observables and entanglement measures.} Beyond one (FM$_x$) and two (AFM$_x$) cavity induced modes, the $\varphi$ setup stabilizes entirely the bond nematic state. In Fig. \ref{fig:Fig3}, we can see the local structure of $\mathcal{Q}^B_{nm}$ and $\mathcal{P}_{nm}$ in the middle of the chain across all the other spins for (a,c) with $J>0$ and (b,d) with $J<0$. This illustrates the structured character of both the magnon pairing and the nematic order in the system with quasi-long range and true long range order,  as confirmed by our quantitative analysis of the quantum correlations. These correlations  present well defined amplitude oscillations across the whole chain similar to frustrated systems~\cite{PhysRevLett.96.027213}.  As nontrivial many body quantum correlations are present in the system, we analyze how entanglement behaves in the system using the entanglement entropy $\mathcal{S}_E=-\mathrm{Tr}[\rho_\Omega\log(\rho_\Omega)]$ considering different subsets $\Omega$ of sites in the chain,  see the insets of Fig.\ref{fig:Fig3}. In contrast to the strength of the $\mathcal{Q}^B_{nm}$, the maximum $\mathcal{S}_E$ is larger for AFM$_z$ coupling as it favors the production of singlets than for the FM$_z$, for all partitions $\Omega$. In (a.1,b.1), we show the entanglement of each of the 3 modes (R,B,G) across the whole lattice induced by the $\varphi$ setting and in (c.1,d.1) the entanglement of the groups of spin singlets embedded in the mode structure. Clearly, the components that form the singlets are more entangled and the singlets themselves have high entanglement, with respect to the G mode. Indeed, this light-matter system acts like a collective entanglement gate protocol naturally and efficiently. In contrast, the entanglement across half the chain is weak, the so called system entanglement [black lines, in (c1,d1)]. Therefore, carefully selecting where to encode information is vital to exploit this entanglement resource. Moreover, achieving the same entanglement properties with a linear local circuit would entail $\mathcal{O}(N^2)$ operations with a prohibitive circuit depth in contrast to our $\mathcal{O}(N)$ cavity controlled scheme, see the SM~\cite{SM} for details. Indeed, the origin of the entanglement properties is completely correlated with the behaviour of the local spin correlations as $\mathcal{Q}^B_{nm}\equiv\langle\{\hat{S}^x_n,\hat{S}^z_m\}\rangle/2$ and the emergence of the long range bond nematic order. Non separability of the quantum state originates due to $\delta\mathcal{Q}^B_{nm}\neq0$. Whenever $\tilde{\mathcal{Q}}^B>0$ and  $\delta\bar{\mathcal{Q}}^B>0$,  we will have cavity induced mode entanglement, which is generated efficiently by $\mathcal{O}(N)$ operations. The behavior across the transition of these light induced modes allows  for a useful qubit resource that in principle could be exploited tuning the light pumped into the system via $V_p$ and  $\Delta_c$. Note that this is only possible due to the fact that the SR modes compete with the  short range Ising background twisting the quantization of the spins due to emergent quadrupolar coupling between different sites across the chain. In the absence of the Ising interaction or the cavity coupling the entanglement is minimal and fragile between the spins.

In conclusion, we have found a path towards an efficient mechanism to generate highly entangled spin structures embedded in spin chains due to cavity light and its combination with short range processes. Moreover, we have identified that this  can be controlled by the geometry of light pumped into the system. It is possible to induce different concentrations of light induced modes by choosing the angle between the cavity axis and the pumped light as $\phi=\arccos(1/(2M-1))$ with $M$ integer,  this choice will generate $M$ modes with non-zero couplings induced by light across the system. The key ingredient to achieve this entanglement resource is the emergence of long range bond nematic order in the system due to the anharmonic response of the spins embedded in the chain. Spin quantum correlations drive the quantum phase transition to the SR state with the emergence of magnon pairs. In principle, these magnon pairs could be used as another source to manipulate information across the chain in finite regions. We have extended our previous developments with Light-matter DMRG and applied the technique to analyze spin systems opening an efficient pathway towards the simulation of these light-matter systems that efficiently captures the quantum correlated nature of spin systems. Notably our work could be used for the optimization entanglement properties in the design of wave guides for analog ion systems. Our methods allow the inclusion of quantum correlated effects beyond mean field schemes in  analog qubit systems of arrays of ions\cite{ions1,ions2}, neutral atoms\cite{Monika,Rey}, Rydberg atoms\cite{CavityRydberg,Rydberg1},  ultracold fermions\cite{JPhillippe} and possibly arrays of giant atoms\cite{Frisk}.

We thank the hospitality of J. Hoffmann of the Physics Department of the University of Gothenburg and the Theoretical Physics Seminar where these ideas were first exposed. We thank H. Ritsch, T. Donner, F. Mivehvar, A. Frisk and A. Bautista-Salvador for useful discussions.
This work is partially supported by the grants UNAM-DGAPA-PAPIIT: IN118823, UNAM-DGPA-PAPIIT: IG101826, UNAM-CIC: Apoyo a Laboratorios Nacionales 2025, CONAHCYT/SECIHITI: LNC-2023-51,  as well as by grants from NVIDIA and utilized NVIDIA RTX 6000 Ada.

\bibliographystyle{apsrev4-2}
\bibliography{main-GDILM_arxiv.bbl}

\begin{thebibliography}{53}%
\makeatletter
\providecommand \@ifxundefined [1]{%
 \@ifx{#1\undefined}
}%
\providecommand \@ifnum [1]{%
 \ifnum #1\expandafter \@firstoftwo
 \else \expandafter \@secondoftwo
 \fi
}%
\providecommand \@ifx [1]{%
 \ifx #1\expandafter \@firstoftwo
 \else \expandafter \@secondoftwo
 \fi
}%
\providecommand \natexlab [1]{#1}%
\providecommand \enquote  [1]{``#1''}%
\providecommand \bibnamefont  [1]{#1}%
\providecommand \bibfnamefont [1]{#1}%
\providecommand \citenamefont [1]{#1}%
\providecommand \href@noop [0]{\@secondoftwo}%
\providecommand \href [0]{\begingroup \@sanitize@url \@href}%
\providecommand \@href[1]{\@@startlink{#1}\@@href}%
\providecommand \@@href[1]{\endgroup#1\@@endlink}%
\providecommand \@sanitize@url [0]{\catcode `\\12\catcode `\$12\catcode
  `\&12\catcode `\#12\catcode `\^12\catcode `\_12\catcode `\%12\relax}%
\providecommand \@@startlink[1]{}%
\providecommand \@@endlink[0]{}%
\providecommand \url  [0]{\begingroup\@sanitize@url \@url }%
\providecommand \@url [1]{\endgroup\@href {#1}{\urlprefix }}%
\providecommand \urlprefix  [0]{URL }%
\providecommand \Eprint [0]{\href }%
\providecommand \doibase [0]{https://doi.org/}%
\providecommand \selectlanguage [0]{\@gobble}%
\providecommand \bibinfo  [0]{\@secondoftwo}%
\providecommand \bibfield  [0]{\@secondoftwo}%
\providecommand \translation [1]{[#1]}%
\providecommand \BibitemOpen [0]{}%
\providecommand \bibitemStop [0]{}%
\providecommand \bibitemNoStop [0]{.\EOS\space}%
\providecommand \EOS [0]{\spacefactor3000\relax}%
\providecommand \BibitemShut  [1]{\csname bibitem#1\endcsname}%
\let\auto@bib@innerbib\@empty
\bibitem [{\citenamefont {Mivehvar}\ \emph {et~al.}(2021)\citenamefont
  {Mivehvar}, \citenamefont {Piazza}, \citenamefont {Donner},\ and\
  \citenamefont {Ritsch}}]{doi:10.1080/00018732.2021.1969727}%
  \BibitemOpen
  \bibfield  {author} {\bibinfo {author} {\bibfnamefont {F.}~\bibnamefont
  {Mivehvar}}, \bibinfo {author} {\bibfnamefont {F.}~\bibnamefont {Piazza}},
  \bibinfo {author} {\bibfnamefont {T.}~\bibnamefont {Donner}},\ and\ \bibinfo
  {author} {\bibfnamefont {H.}~\bibnamefont {Ritsch}},\ }\bibfield  {title}
  {\bibinfo {title} {{Cavity QED with quantum gases: new paradigms in many-body
  physics}},\ }\href {https://doi.org/10.1080/00018732.2021.1969727} {\bibfield
   {journal} {\bibinfo  {journal} {Advances in Physics}\ }\textbf {\bibinfo
  {volume} {70}},\ \bibinfo {pages} {1} (\bibinfo {year} {2021})}\BibitemShut
  {NoStop}%
\bibitem [{\citenamefont {Elliott}\ \emph {et~al.}(2015)\citenamefont
  {Elliott}, \citenamefont {Kozlowski}, \citenamefont {Caballero-Benitez},\
  and\ \citenamefont {Mekhov}}]{PhysRevLett.114.113604}%
  \BibitemOpen
  \bibfield  {author} {\bibinfo {author} {\bibfnamefont {T.~J.}\ \bibnamefont
  {Elliott}}, \bibinfo {author} {\bibfnamefont {W.}~\bibnamefont {Kozlowski}},
  \bibinfo {author} {\bibfnamefont {S.~F.}\ \bibnamefont {Caballero-Benitez}},\
  and\ \bibinfo {author} {\bibfnamefont {I.~B.}\ \bibnamefont {Mekhov}},\
  }\bibfield  {title} {\bibinfo {title} {{Multipartite Entangled Spatial Modes
  of Ultracold Atoms Generated and Controlled by Quantum Measurement}},\ }\href
  {https://doi.org/10.1103/PhysRevLett.114.113604} {\bibfield  {journal}
  {\bibinfo  {journal} {Phys. Rev. Lett.}\ }\textbf {\bibinfo {volume} {114}},\
  \bibinfo {pages} {113604} (\bibinfo {year} {2015})}\BibitemShut {NoStop}%
\bibitem [{\citenamefont {Mazzucchi}\ \emph
  {et~al.}(2016{\natexlab{a}})\citenamefont {Mazzucchi}, \citenamefont
  {Kozlowski}, \citenamefont {Caballero-Benitez}, \citenamefont {Elliott},\
  and\ \citenamefont {Mekhov}}]{PhysRevA.93.023632}%
  \BibitemOpen
  \bibfield  {author} {\bibinfo {author} {\bibfnamefont {G.}~\bibnamefont
  {Mazzucchi}}, \bibinfo {author} {\bibfnamefont {W.}~\bibnamefont
  {Kozlowski}}, \bibinfo {author} {\bibfnamefont {S.~F.}\ \bibnamefont
  {Caballero-Benitez}}, \bibinfo {author} {\bibfnamefont {T.~J.}\ \bibnamefont
  {Elliott}},\ and\ \bibinfo {author} {\bibfnamefont {I.~B.}\ \bibnamefont
  {Mekhov}},\ }\bibfield  {title} {\bibinfo {title} {Quantum
  measurement-induced dynamics of many-body ultracold bosonic and fermionic
  systems in optical lattices},\ }\href
  {https://doi.org/10.1103/PhysRevA.93.023632} {\bibfield  {journal} {\bibinfo
  {journal} {Phys. Rev. A}\ }\textbf {\bibinfo {volume} {93}},\ \bibinfo
  {pages} {023632} (\bibinfo {year} {2016}{\natexlab{a}})}\BibitemShut
  {NoStop}%
\bibitem [{\citenamefont {Caballero-Benitez}\ and\ \citenamefont
  {Mekhov}(2015{\natexlab{a}})}]{PhysRevLett.115.243604}%
  \BibitemOpen
  \bibfield  {author} {\bibinfo {author} {\bibfnamefont {S.~F.}\ \bibnamefont
  {Caballero-Benitez}}\ and\ \bibinfo {author} {\bibfnamefont {I.~B.}\
  \bibnamefont {Mekhov}},\ }\bibfield  {title} {\bibinfo {title} {{Quantum
  Optical Lattices for Emergent Many-Body Phases of Ultracold Atoms}},\ }\href
  {https://doi.org/10.1103/PhysRevLett.115.243604} {\bibfield  {journal}
  {\bibinfo  {journal} {Phys. Rev. Lett.}\ }\textbf {\bibinfo {volume} {115}},\
  \bibinfo {pages} {243604} (\bibinfo {year} {2015}{\natexlab{a}})}\BibitemShut
  {NoStop}%
\bibitem [{\citenamefont {Baumann}\ \emph {et~al.}(2010)\citenamefont
  {Baumann}, \citenamefont {Guerlin}, \citenamefont {Brennecke},\ and\
  \citenamefont {Esslinger}}]{Baumann2010}%
  \BibitemOpen
  \bibfield  {author} {\bibinfo {author} {\bibfnamefont {K.}~\bibnamefont
  {Baumann}}, \bibinfo {author} {\bibfnamefont {C.}~\bibnamefont {Guerlin}},
  \bibinfo {author} {\bibfnamefont {F.}~\bibnamefont {Brennecke}},\ and\
  \bibinfo {author} {\bibfnamefont {T.}~\bibnamefont {Esslinger}},\ }\bibfield
  {title} {\bibinfo {title} {{Dicke quantum phase transition with a superfluid
  gas in an optical cavity}},\ }\href {https://doi.org/10.1038/nature09009}
  {\bibfield  {journal} {\bibinfo  {journal} {Nature}\ }\textbf {\bibinfo
  {volume} {464}},\ \bibinfo {pages} {1301} (\bibinfo {year}
  {2010})}\BibitemShut {NoStop}%
\bibitem [{\citenamefont {Li}\ \emph {et~al.}(2021)\citenamefont {Li},
  \citenamefont {Dreon}, \citenamefont {Zupancic}, \citenamefont
  {Baumg\"artner}, \citenamefont {Morales}, \citenamefont {Zheng},
  \citenamefont {Cooper}, \citenamefont {Donner},\ and\ \citenamefont
  {Esslinger}}]{PhysRevResearch.3.L012024}%
  \BibitemOpen
  \bibfield  {author} {\bibinfo {author} {\bibfnamefont {X.}~\bibnamefont
  {Li}}, \bibinfo {author} {\bibfnamefont {D.}~\bibnamefont {Dreon}}, \bibinfo
  {author} {\bibfnamefont {P.}~\bibnamefont {Zupancic}}, \bibinfo {author}
  {\bibfnamefont {A.}~\bibnamefont {Baumg\"artner}}, \bibinfo {author}
  {\bibfnamefont {A.}~\bibnamefont {Morales}}, \bibinfo {author} {\bibfnamefont
  {W.}~\bibnamefont {Zheng}}, \bibinfo {author} {\bibfnamefont {N.~R.}\
  \bibnamefont {Cooper}}, \bibinfo {author} {\bibfnamefont {T.}~\bibnamefont
  {Donner}},\ and\ \bibinfo {author} {\bibfnamefont {T.}~\bibnamefont
  {Esslinger}},\ }\bibfield  {title} {\bibinfo {title} {First order phase
  transition between two centro-symmetric superradiant crystals},\ }\href
  {https://doi.org/10.1103/PhysRevResearch.3.L012024} {\bibfield  {journal}
  {\bibinfo  {journal} {Phys. Rev. Res.}\ }\textbf {\bibinfo {volume} {3}},\
  \bibinfo {pages} {L012024} (\bibinfo {year} {2021})}\BibitemShut {NoStop}%
\bibitem [{\citenamefont {Sachdev}(2011)}]{Sachdev2011}%
  \BibitemOpen
  \bibfield  {author} {\bibinfo {author} {\bibfnamefont {S.}~\bibnamefont
  {Sachdev}},\ }\href {https://doi.org/10.1017/CBO9780511973765} {\emph
  {\bibinfo {title} {{Quantum Phase Transitions}}}},\ \bibinfo {edition} {2nd}\
  ed.\ (\bibinfo  {publisher} {Cambridge University Press},\ \bibinfo {address}
  {Cambridge},\ \bibinfo {year} {2011})\BibitemShut {NoStop}%
\bibitem [{\citenamefont {Landig}\ \emph {et~al.}(2016)\citenamefont {Landig},
  \citenamefont {Hruby}, \citenamefont {Dogra}, \citenamefont {Landini},
  \citenamefont {Mottl}, \citenamefont {Donner},\ and\ \citenamefont
  {Esslinger}}]{Landig2016}%
  \BibitemOpen
  \bibfield  {author} {\bibinfo {author} {\bibfnamefont {R.}~\bibnamefont
  {Landig}}, \bibinfo {author} {\bibfnamefont {L.}~\bibnamefont {Hruby}},
  \bibinfo {author} {\bibfnamefont {N.}~\bibnamefont {Dogra}}, \bibinfo
  {author} {\bibfnamefont {M.}~\bibnamefont {Landini}}, \bibinfo {author}
  {\bibfnamefont {R.}~\bibnamefont {Mottl}}, \bibinfo {author} {\bibfnamefont
  {T.}~\bibnamefont {Donner}},\ and\ \bibinfo {author} {\bibfnamefont
  {T.}~\bibnamefont {Esslinger}},\ }\bibfield  {title} {\bibinfo {title}
  {Quantum phases from competing short- and long-range interactions in an
  optical lattice},\ }\href {https://doi.org/10.1038/nature17409} {\bibfield
  {journal} {\bibinfo  {journal} {Nature}\ }\textbf {\bibinfo {volume} {532}},\
  \bibinfo {pages} {476} (\bibinfo {year} {2016})}\BibitemShut {NoStop}%
\bibitem [{\citenamefont {Klinder}\ \emph {et~al.}(2015)\citenamefont
  {Klinder}, \citenamefont {Ke\ss~ler}, \citenamefont {Reza~Bakhtiari},
  \citenamefont {Thorwart},\ and\ \citenamefont {Hemmerich}}]{Hemmerich}%
  \BibitemOpen
  \bibfield  {author} {\bibinfo {author} {\bibfnamefont {J.}~\bibnamefont
  {Klinder}}, \bibinfo {author} {\bibfnamefont {H.}~\bibnamefont {Ke\ss~ler}},
  \bibinfo {author} {\bibfnamefont {M.}~\bibnamefont {Reza~Bakhtiari}},
  \bibinfo {author} {\bibfnamefont {M.}~\bibnamefont {Thorwart}},\ and\
  \bibinfo {author} {\bibfnamefont {A.}~\bibnamefont {Hemmerich}},\ }\bibfield
  {title} {\bibinfo {title} {{Observation of a Superradiant Mott Insulator in
  the Dicke-Hubbard Models}},\ }\href
  {https://doi.org/10.1103/PhysRevLett.115.230403} {\bibfield  {journal}
  {\bibinfo  {journal} {Phys. Rev. Lett.}\ }\textbf {\bibinfo {volume} {115}},\
  \bibinfo {pages} {230403} (\bibinfo {year} {2015})}\BibitemShut {NoStop}%
\bibitem [{\citenamefont {L{\'e}onard}\ \emph {et~al.}(2017)\citenamefont
  {L{\'e}onard}, \citenamefont {Morales}, \citenamefont {Zupancic},
  \citenamefont {Esslinger},\ and\ \citenamefont {Donner}}]{Leonard2017}%
  \BibitemOpen
  \bibfield  {author} {\bibinfo {author} {\bibfnamefont {J.}~\bibnamefont
  {L{\'e}onard}}, \bibinfo {author} {\bibfnamefont {A.}~\bibnamefont
  {Morales}}, \bibinfo {author} {\bibfnamefont {P.}~\bibnamefont {Zupancic}},
  \bibinfo {author} {\bibfnamefont {T.}~\bibnamefont {Esslinger}},\ and\
  \bibinfo {author} {\bibfnamefont {T.}~\bibnamefont {Donner}},\ }\bibfield
  {title} {\bibinfo {title} {Supersolid formation in a quantum gas breaking a
  continuous translational symmetry},\ }\href
  {https://doi.org/10.1038/nature21067} {\bibfield  {journal} {\bibinfo
  {journal} {Nature}\ }\textbf {\bibinfo {volume} {543}},\ \bibinfo {pages}
  {87} (\bibinfo {year} {2017})}\BibitemShut {NoStop}%
\bibitem [{\citenamefont {Kroeze}\ \emph {et~al.}(2018)\citenamefont {Kroeze},
  \citenamefont {Guo}, \citenamefont {Vaidya}, \citenamefont {Keeling},\ and\
  \citenamefont {Lev}}]{PhysRevLett.121.163601}%
  \BibitemOpen
  \bibfield  {author} {\bibinfo {author} {\bibfnamefont {R.~M.}\ \bibnamefont
  {Kroeze}}, \bibinfo {author} {\bibfnamefont {Y.}~\bibnamefont {Guo}},
  \bibinfo {author} {\bibfnamefont {V.~D.}\ \bibnamefont {Vaidya}}, \bibinfo
  {author} {\bibfnamefont {J.}~\bibnamefont {Keeling}},\ and\ \bibinfo {author}
  {\bibfnamefont {B.~L.}\ \bibnamefont {Lev}},\ }\bibfield  {title} {\bibinfo
  {title} {{Spinor Self-Ordering of a Quantum Gas in a Cavity}},\ }\href
  {https://doi.org/10.1103/PhysRevLett.121.163601} {\bibfield  {journal}
  {\bibinfo  {journal} {Phys. Rev. Lett.}\ }\textbf {\bibinfo {volume} {121}},\
  \bibinfo {pages} {163601} (\bibinfo {year} {2018})}\BibitemShut {NoStop}%
\bibitem [{\citenamefont {Lozano-M\'endez}\ \emph {et~al.}(2022)\citenamefont
  {Lozano-M\'endez}, \citenamefont {C\'asares},\ and\ \citenamefont
  {Caballero-Ben\'{\i}tez}}]{PhysRevLett.128.080601}%
  \BibitemOpen
  \bibfield  {author} {\bibinfo {author} {\bibfnamefont {K.}~\bibnamefont
  {Lozano-M\'endez}}, \bibinfo {author} {\bibfnamefont {A.~H.}\ \bibnamefont
  {C\'asares}},\ and\ \bibinfo {author} {\bibfnamefont {S.~F.}\ \bibnamefont
  {Caballero-Ben\'{\i}tez}},\ }\bibfield  {title} {\bibinfo {title} {{Spin
  Entanglement and Magnetic Competition via Long-Range Interactions in Spinor
  Quantum Optical Lattices}},\ }\href
  {https://doi.org/10.1103/PhysRevLett.128.080601} {\bibfield  {journal}
  {\bibinfo  {journal} {Phys. Rev. Lett.}\ }\textbf {\bibinfo {volume} {128}},\
  \bibinfo {pages} {080601} (\bibinfo {year} {2022})}\BibitemShut {NoStop}%
\bibitem [{\citenamefont {R\'ios-S\'anchez}\ and\ \citenamefont
  {Caballero-Ben\'itez}(2025)}]{Brahyam}%
  \BibitemOpen
  \bibfield  {author} {\bibinfo {author} {\bibfnamefont {B.}~\bibnamefont
  {R\'ios-S\'anchez}}\ and\ \bibinfo {author} {\bibfnamefont {S.~F.}\
  \bibnamefont {Caballero-Ben\'itez}},\ }\bibfield  {title} {\bibinfo {title}
  {{Control, competition and coexistence of effective magnetic orders by
  interactions in Bose-Einstein condensates with high-Q cavities}},\ }\href
  {https://doi.org/10.1088/1367-2630/ae1b32} {\bibfield  {journal} {\bibinfo
  {journal} {New J. Phys.}\ }\textbf {\bibinfo {volume} {27}},\ \bibinfo
  {pages} {113203} (\bibinfo {year} {2025})}\BibitemShut {NoStop}%
\bibitem [{\citenamefont {Song}\ \emph {et~al.}(2025)\citenamefont {Song},
  \citenamefont {Barberena}, \citenamefont {Young}, \citenamefont {Chaparro},
  \citenamefont {Chu}, \citenamefont {Agarwal}, \citenamefont {Niu},
  \citenamefont {Young}, \citenamefont {Rey},\ and\ \citenamefont
  {Thompson}}]{Rey}%
  \BibitemOpen
  \bibfield  {author} {\bibinfo {author} {\bibfnamefont {E.~Y.}\ \bibnamefont
  {Song}}, \bibinfo {author} {\bibfnamefont {D.}~\bibnamefont {Barberena}},
  \bibinfo {author} {\bibfnamefont {D.~J.}\ \bibnamefont {Young}}, \bibinfo
  {author} {\bibfnamefont {E.}~\bibnamefont {Chaparro}}, \bibinfo {author}
  {\bibfnamefont {A.}~\bibnamefont {Chu}}, \bibinfo {author} {\bibfnamefont
  {S.}~\bibnamefont {Agarwal}}, \bibinfo {author} {\bibfnamefont
  {Z.}~\bibnamefont {Niu}}, \bibinfo {author} {\bibfnamefont {J.~T.}\
  \bibnamefont {Young}}, \bibinfo {author} {\bibfnamefont {A.~M.}\ \bibnamefont
  {Rey}},\ and\ \bibinfo {author} {\bibfnamefont {J.~K.}\ \bibnamefont
  {Thompson}},\ }\bibfield  {title} {\bibinfo {title} {{A dissipation-induced
  superradiant transition in a strontium cavity-QED system}},\ }\href
  {https://doi.org/10.1126/sciadv.adu5799} {\bibfield  {journal} {\bibinfo
  {journal} {Science Advances}\ }\textbf {\bibinfo {volume} {11}},\ \bibinfo
  {pages} {eadu5799} (\bibinfo {year} {2025})},\ \Eprint
  {https://arxiv.org/abs/https://www.science.org/doi/pdf/10.1126/sciadv.adu5799}
  {https://www.science.org/doi/pdf/10.1126/sciadv.adu5799} \BibitemShut
  {NoStop}%
\bibitem [{\citenamefont {Mamaev}\ \emph {et~al.}(2021)\citenamefont {Mamaev},
  \citenamefont {Kimchi}, \citenamefont {Nandkishore},\ and\ \citenamefont
  {Rey}}]{PhysRevResearch.3.013178}%
  \BibitemOpen
  \bibfield  {author} {\bibinfo {author} {\bibfnamefont {M.}~\bibnamefont
  {Mamaev}}, \bibinfo {author} {\bibfnamefont {I.}~\bibnamefont {Kimchi}},
  \bibinfo {author} {\bibfnamefont {R.~M.}\ \bibnamefont {Nandkishore}},\ and\
  \bibinfo {author} {\bibfnamefont {A.~M.}\ \bibnamefont {Rey}},\ }\bibfield
  {title} {\bibinfo {title} {Tunable-spin-model generation with
  spin-orbit-coupled fermions in optical lattices},\ }\href
  {https://doi.org/10.1103/PhysRevResearch.3.013178} {\bibfield  {journal}
  {\bibinfo  {journal} {Phys. Rev. Res.}\ }\textbf {\bibinfo {volume} {3}},\
  \bibinfo {pages} {013178} (\bibinfo {year} {2021})}\BibitemShut {NoStop}%
\bibitem [{\citenamefont {Yasir}\ and\ \citenamefont
  {Liu}(2026)}]{CavityRydberg}%
  \BibitemOpen
  \bibfield  {author} {\bibinfo {author} {\bibfnamefont {K.~A.}\ \bibnamefont
  {Yasir}}\ and\ \bibinfo {author} {\bibfnamefont {W.-M.}\ \bibnamefont
  {Liu}},\ }\href@noop {} {\emph {\bibinfo {title} {{ Rydberg Atoms in
  Cavity}}}},\ \bibinfo {edition} {1st}\ ed.\ (\bibinfo  {publisher} {Springer
  Nature},\ \bibinfo {address} {Singapore},\ \bibinfo {year}
  {2026})\BibitemShut {NoStop}%
\bibitem [{\citenamefont {De~Santis}\ \emph {et~al.}(2026)\citenamefont
  {De~Santis}, \citenamefont {Dura-Kov\'acs}, \citenamefont {\"Onc\"u},
  \citenamefont {Bouscal}, \citenamefont {Vasileiadis},\ and\ \citenamefont
  {Zeiher}}]{Rydberg1}%
  \BibitemOpen
  \bibfield  {author} {\bibinfo {author} {\bibfnamefont {J.}~\bibnamefont
  {De~Santis}}, \bibinfo {author} {\bibfnamefont {B.}~\bibnamefont
  {Dura-Kov\'acs}}, \bibinfo {author} {\bibfnamefont {M.}~\bibnamefont
  {\"Onc\"u}}, \bibinfo {author} {\bibfnamefont {A.}~\bibnamefont {Bouscal}},
  \bibinfo {author} {\bibfnamefont {D.}~\bibnamefont {Vasileiadis}},\ and\
  \bibinfo {author} {\bibfnamefont {J.}~\bibnamefont {Zeiher}},\ }\href
  {https://doi.org/10.48550/arXiv.2602.12152} {\bibinfo {title} {{Realization
  of a cavity-coupled Rydberg array}}} (\bibinfo {year} {2026}),\ \Eprint
  {https://arxiv.org/abs/2602.12152} {arXiv:2602.12152 [quant-ph]} \BibitemShut
  {NoStop}%
\bibitem [{\citenamefont {Gelhausen}\ \emph {et~al.}(2016)\citenamefont
  {Gelhausen}, \citenamefont {Buchhold}, \citenamefont {Rosch},\ and\
  \citenamefont {Strack}}]{SciPostPhys.1.1.004}%
  \BibitemOpen
  \bibfield  {author} {\bibinfo {author} {\bibfnamefont {J.}~\bibnamefont
  {Gelhausen}}, \bibinfo {author} {\bibfnamefont {M.}~\bibnamefont {Buchhold}},
  \bibinfo {author} {\bibfnamefont {A.}~\bibnamefont {Rosch}},\ and\ \bibinfo
  {author} {\bibfnamefont {P.}~\bibnamefont {Strack}},\ }\bibfield  {title}
  {\bibinfo {title} {{Quantum-optical magnets with competing short- and
  long-range interactions: Rydberg-dressed spin lattice in an optical
  cavity}},\ }\href {https://doi.org/10.21468/SciPostPhys.1.1.004} {\bibfield
  {journal} {\bibinfo  {journal} {SciPost Phys.}\ }\textbf {\bibinfo {volume}
  {1}},\ \bibinfo {pages} {004} (\bibinfo {year} {2016})}\BibitemShut {NoStop}%
\bibitem [{\citenamefont {Srinivas}\ \emph {et~al.}(2019)\citenamefont
  {Srinivas}, \citenamefont {Burd}, \citenamefont {Sutherland}, \citenamefont
  {Wilson}, \citenamefont {Wineland}, \citenamefont {Leibfried}, \citenamefont
  {Allcock},\ and\ \citenamefont {Slichter}}]{ions1}%
  \BibitemOpen
  \bibfield  {author} {\bibinfo {author} {\bibfnamefont {R.}~\bibnamefont
  {Srinivas}}, \bibinfo {author} {\bibfnamefont {S.~C.}\ \bibnamefont {Burd}},
  \bibinfo {author} {\bibfnamefont {R.~T.}\ \bibnamefont {Sutherland}},
  \bibinfo {author} {\bibfnamefont {A.~C.}\ \bibnamefont {Wilson}}, \bibinfo
  {author} {\bibfnamefont {D.~J.}\ \bibnamefont {Wineland}}, \bibinfo {author}
  {\bibfnamefont {D.}~\bibnamefont {Leibfried}}, \bibinfo {author}
  {\bibfnamefont {D.~T.~C.}\ \bibnamefont {Allcock}},\ and\ \bibinfo {author}
  {\bibfnamefont {D.~H.}\ \bibnamefont {Slichter}},\ }\bibfield  {title}
  {\bibinfo {title} {{Trapped-Ion Spin-Motion Coupling with Microwaves and a
  Near-Motional Oscillating Magnetic Field Gradient}},\ }\href
  {https://doi.org/10.1103/PhysRevLett.122.163201} {\bibfield  {journal}
  {\bibinfo  {journal} {Phys. Rev. Lett.}\ }\textbf {\bibinfo {volume} {122}},\
  \bibinfo {pages} {163201} (\bibinfo {year} {2019})}\BibitemShut {NoStop}%
\bibitem [{\citenamefont {Zarantonello}\ \emph {et~al.}(2019)\citenamefont
  {Zarantonello}, \citenamefont {Hahn}, \citenamefont {Morgner}, \citenamefont
  {Schulte}, \citenamefont {Bautista-Salvador}, \citenamefont {Werner},
  \citenamefont {Hammerer},\ and\ \citenamefont {Ospelkaus}}]{ions2}%
  \BibitemOpen
  \bibfield  {author} {\bibinfo {author} {\bibfnamefont {G.}~\bibnamefont
  {Zarantonello}}, \bibinfo {author} {\bibfnamefont {H.}~\bibnamefont {Hahn}},
  \bibinfo {author} {\bibfnamefont {J.}~\bibnamefont {Morgner}}, \bibinfo
  {author} {\bibfnamefont {M.}~\bibnamefont {Schulte}}, \bibinfo {author}
  {\bibfnamefont {A.}~\bibnamefont {Bautista-Salvador}}, \bibinfo {author}
  {\bibfnamefont {R.~F.}\ \bibnamefont {Werner}}, \bibinfo {author}
  {\bibfnamefont {K.}~\bibnamefont {Hammerer}},\ and\ \bibinfo {author}
  {\bibfnamefont {C.}~\bibnamefont {Ospelkaus}},\ }\bibfield  {title} {\bibinfo
  {title} {{Robust and Resource-Efficient Microwave Near-Field Entangling
  $^{9}{\mathrm{Be}}^{+}$ Gate}},\ }\href
  {https://doi.org/10.1103/PhysRevLett.123.260503} {\bibfield  {journal}
  {\bibinfo  {journal} {Phys. Rev. Lett.}\ }\textbf {\bibinfo {volume} {123}},\
  \bibinfo {pages} {260503} (\bibinfo {year} {2019})}\BibitemShut {NoStop}%
\bibitem [{\citenamefont {Alvarez-Giron}\ \emph {et~al.}(2024)\citenamefont
  {Alvarez-Giron}, \citenamefont {Solano}, \citenamefont {Sinha},\ and\
  \citenamefont {Barberis-Blostein}}]{PhysRevResearch.6.023213}%
  \BibitemOpen
  \bibfield  {author} {\bibinfo {author} {\bibfnamefont {W.}~\bibnamefont
  {Alvarez-Giron}}, \bibinfo {author} {\bibfnamefont {P.}~\bibnamefont
  {Solano}}, \bibinfo {author} {\bibfnamefont {K.}~\bibnamefont {Sinha}},\ and\
  \bibinfo {author} {\bibfnamefont {P.}~\bibnamefont {Barberis-Blostein}},\
  }\bibfield  {title} {\bibinfo {title} {Delay-induced spontaneous dark-state
  generation from two distant excited atoms},\ }\href
  {https://doi.org/10.1103/PhysRevResearch.6.023213} {\bibfield  {journal}
  {\bibinfo  {journal} {Phys. Rev. Res.}\ }\textbf {\bibinfo {volume} {6}},\
  \bibinfo {pages} {023213} (\bibinfo {year} {2024})}\BibitemShut {NoStop}%
\bibitem [{\citenamefont {Solano}\ \emph {et~al.}(2017)\citenamefont {Solano},
  \citenamefont {Barberis-Blostein}, \citenamefont {Fatemi}, \citenamefont
  {Orozco},\ and\ \citenamefont {Rolston}}]{Solano2017}%
  \BibitemOpen
  \bibfield  {author} {\bibinfo {author} {\bibfnamefont {P.}~\bibnamefont
  {Solano}}, \bibinfo {author} {\bibfnamefont {P.}~\bibnamefont
  {Barberis-Blostein}}, \bibinfo {author} {\bibfnamefont {F.~K.}\ \bibnamefont
  {Fatemi}}, \bibinfo {author} {\bibfnamefont {L.~A.}\ \bibnamefont {Orozco}},\
  and\ \bibinfo {author} {\bibfnamefont {S.~L.}\ \bibnamefont {Rolston}},\
  }\bibfield  {title} {\bibinfo {title} {Super-radiance reveals infinite-range
  dipole interactions through a nanofiber},\ }\href
  {https://doi.org/10.1038/s41467-017-01994-3} {\bibfield  {journal} {\bibinfo
  {journal} {Nature Communications}\ }\textbf {\bibinfo {volume} {8}},\
  \bibinfo {pages} {1857} (\bibinfo {year} {2017})}\BibitemShut {NoStop}%
\bibitem [{\citenamefont {Ram\'{\i}rez-Barajas}\ and\ \citenamefont
  {Caballero-Benitez}(2025)}]{gvm2-b46t}%
  \BibitemOpen
  \bibfield  {author} {\bibinfo {author} {\bibfnamefont {A.~U.}\ \bibnamefont
  {Ram\'{\i}rez-Barajas}}\ and\ \bibinfo {author} {\bibfnamefont {S.~F.}\
  \bibnamefont {Caballero-Benitez}},\ }\bibfield  {title} {\bibinfo {title}
  {{Structural Dynamics and Strong Correlations in Dynamical Quantum Optical
  Lattices}},\ }\href {https://doi.org/10.1103/gvm2-b46t} {\bibfield  {journal}
  {\bibinfo  {journal} {Phys. Rev. Lett.}\ }\textbf {\bibinfo {volume} {135}},\
  \bibinfo {pages} {120602} (\bibinfo {year} {2025})}\BibitemShut {NoStop}%
\bibitem [{SM()}]{SM}%
  \BibitemOpen
  \href@noop {} {}\bibinfo {note} {See Supplemental Material at [URL] for
  cavity light details, computational methods, finite size scaling analysis,
  local gate circuit and additional details of the order parameters, which
  includes Refs. [39-53]}\BibitemShut {NoStop}%
\bibitem [{\citenamefont {Leibig}\ \emph {et~al.}(2026)\citenamefont {Leibig},
  \citenamefont {H\"ormann}, \citenamefont {Langheld}, \citenamefont
  {Schellenberger},\ and\ \citenamefont {Schmidt}}]{DIp}%
  \BibitemOpen
  \bibfield  {author} {\bibinfo {author} {\bibfnamefont {J.}~\bibnamefont
  {Leibig}}, \bibinfo {author} {\bibfnamefont {M.}~\bibnamefont {H\"ormann}},
  \bibinfo {author} {\bibfnamefont {A.}~\bibnamefont {Langheld}}, \bibinfo
  {author} {\bibfnamefont {A.}~\bibnamefont {Schellenberger}},\ and\ \bibinfo
  {author} {\bibfnamefont {K.}~\bibnamefont {Schmidt}},\ }\href
  {https://doi.org/10.48550/arXiv.2601.10210} {\bibinfo {title} {{Quantitative
  approach for the Dicke-Ising chain with an effective self-consistent matter
  Hamiltonian}}} (\bibinfo {year} {2026}),\ \Eprint
  {https://arxiv.org/abs/2601.10210} {arXiv:2601.10210 [quant-ph]} \BibitemShut
  {NoStop}%
\bibitem [{\citenamefont {Langheld}\ \emph {et~al.}(2025)\citenamefont
  {Langheld}, \citenamefont {H\"ormann},\ and\ \citenamefont
  {Schmidt}}]{lcvj-ksct}%
  \BibitemOpen
  \bibfield  {author} {\bibinfo {author} {\bibfnamefont {A.}~\bibnamefont
  {Langheld}}, \bibinfo {author} {\bibfnamefont {M.}~\bibnamefont
  {H\"ormann}},\ and\ \bibinfo {author} {\bibfnamefont {K.~P.}\ \bibnamefont
  {Schmidt}},\ }\bibfield  {title} {\bibinfo {title} {{Quantum phase diagrams
  of Dicke-Ising models by a wormhole algorithm}},\ }\href
  {https://doi.org/10.1103/lcvj-ksct} {\bibfield  {journal} {\bibinfo
  {journal} {Phys. Rev. B}\ }\textbf {\bibinfo {volume} {112}},\ \bibinfo
  {pages} {L161123} (\bibinfo {year} {2025})}\BibitemShut {NoStop}%
\bibitem [{\citenamefont {Rohn}\ \emph {et~al.}(2020)\citenamefont {Rohn},
  \citenamefont {H\"ormann}, \citenamefont {Genes},\ and\ \citenamefont
  {Schmidt}}]{PhysRevResearch.2.023131}%
  \BibitemOpen
  \bibfield  {author} {\bibinfo {author} {\bibfnamefont {J.}~\bibnamefont
  {Rohn}}, \bibinfo {author} {\bibfnamefont {M.}~\bibnamefont {H\"ormann}},
  \bibinfo {author} {\bibfnamefont {C.}~\bibnamefont {Genes}},\ and\ \bibinfo
  {author} {\bibfnamefont {K.~P.}\ \bibnamefont {Schmidt}},\ }\bibfield
  {title} {\bibinfo {title} {Ising model in a light-induced quantized
  transverse field},\ }\href {https://doi.org/10.1103/PhysRevResearch.2.023131}
  {\bibfield  {journal} {\bibinfo  {journal} {Phys. Rev. Res.}\ }\textbf
  {\bibinfo {volume} {2}},\ \bibinfo {pages} {023131} (\bibinfo {year}
  {2020})}\BibitemShut {NoStop}%
\bibitem [{\citenamefont {Maschler}\ \emph {et~al.}(2008)\citenamefont
  {Maschler}, \citenamefont {Mekhov},\ and\ \citenamefont
  {Ritsch}}]{Maschler2008}%
  \BibitemOpen
  \bibfield  {author} {\bibinfo {author} {\bibfnamefont {C.}~\bibnamefont
  {Maschler}}, \bibinfo {author} {\bibfnamefont {I.~B.}\ \bibnamefont
  {Mekhov}},\ and\ \bibinfo {author} {\bibfnamefont {H.}~\bibnamefont
  {Ritsch}},\ }\bibfield  {title} {\bibinfo {title} {{Ultracold atoms in
  optical lattices generated by quantized light fields}},\ }\href
  {https://doi.org/10.1140/epjd/e2008-00016-4} {\bibfield  {journal} {\bibinfo
  {journal} {The European Physical Journal D}\ }\textbf {\bibinfo {volume}
  {46}},\ \bibinfo {pages} {545} (\bibinfo {year} {2008})}\BibitemShut
  {NoStop}%
\bibitem [{\citenamefont {Caballero-Benitez}\ and\ \citenamefont
  {Mekhov}(2015{\natexlab{b}})}]{NewJPhys.17.123023}%
  \BibitemOpen
  \bibfield  {author} {\bibinfo {author} {\bibfnamefont {S.~F.}\ \bibnamefont
  {Caballero-Benitez}}\ and\ \bibinfo {author} {\bibfnamefont {I.~B.}\
  \bibnamefont {Mekhov}},\ }\bibfield  {title} {\bibinfo {title} {Quantum
  properties of light scattered from structured many-body phases of ultracold
  atoms in quantum optical lattices},\ }\href
  {https://doi.org/10.1088/1367-2630/17/12/123023} {\bibfield  {journal}
  {\bibinfo  {journal} {New J. Phys}\ }\textbf {\bibinfo {volume} {17}},\
  \bibinfo {pages} {123023} (\bibinfo {year} {2015}{\natexlab{b}})}\BibitemShut
  {NoStop}%
\bibitem [{\citenamefont {Morales}\ \emph {et~al.}(2019)\citenamefont
  {Morales}, \citenamefont {Dreon}, \citenamefont {Li}, \citenamefont
  {Baumg\"artner}, \citenamefont {Zupancic}, \citenamefont {Donner},\ and\
  \citenamefont {Esslinger}}]{PhysRevA.100.013816}%
  \BibitemOpen
  \bibfield  {author} {\bibinfo {author} {\bibfnamefont {A.}~\bibnamefont
  {Morales}}, \bibinfo {author} {\bibfnamefont {D.}~\bibnamefont {Dreon}},
  \bibinfo {author} {\bibfnamefont {X.}~\bibnamefont {Li}}, \bibinfo {author}
  {\bibfnamefont {A.}~\bibnamefont {Baumg\"artner}}, \bibinfo {author}
  {\bibfnamefont {P.}~\bibnamefont {Zupancic}}, \bibinfo {author}
  {\bibfnamefont {T.}~\bibnamefont {Donner}},\ and\ \bibinfo {author}
  {\bibfnamefont {T.}~\bibnamefont {Esslinger}},\ }\bibfield  {title} {\bibinfo
  {title} {{Two-mode Dicke model from nondegenerate polarization modes}},\
  }\href {https://doi.org/10.1103/PhysRevA.100.013816} {\bibfield  {journal}
  {\bibinfo  {journal} {Phys. Rev. A}\ }\textbf {\bibinfo {volume} {100}},\
  \bibinfo {pages} {013816} (\bibinfo {year} {2019})}\BibitemShut {NoStop}%
\bibitem [{\citenamefont {Shannon}\ \emph {et~al.}(2006)\citenamefont
  {Shannon}, \citenamefont {Momoi},\ and\ \citenamefont
  {Sindzingre}}]{PhysRevLett.96.027213}%
  \BibitemOpen
  \bibfield  {author} {\bibinfo {author} {\bibfnamefont {N.}~\bibnamefont
  {Shannon}}, \bibinfo {author} {\bibfnamefont {T.}~\bibnamefont {Momoi}},\
  and\ \bibinfo {author} {\bibfnamefont {P.}~\bibnamefont {Sindzingre}},\
  }\bibfield  {title} {\bibinfo {title} {{Nematic Order in Square Lattice
  Frustrated Ferromagnets}},\ }\href
  {https://doi.org/10.1103/PhysRevLett.96.027213} {\bibfield  {journal}
  {\bibinfo  {journal} {Phys. Rev. Lett.}\ }\textbf {\bibinfo {volume} {96}},\
  \bibinfo {pages} {027213} (\bibinfo {year} {2006})}\BibitemShut {NoStop}%
\bibitem [{\citenamefont {Jiang}\ \emph {et~al.}(2023)\citenamefont {Jiang},
  \citenamefont {Romh\'anyi}, \citenamefont {White}, \citenamefont
  {Zhitomirsky},\ and\ \citenamefont {Chernyshev}}]{PhysRevLett.130.116701}%
  \BibitemOpen
  \bibfield  {author} {\bibinfo {author} {\bibfnamefont {S.}~\bibnamefont
  {Jiang}}, \bibinfo {author} {\bibfnamefont {J.}~\bibnamefont {Romh\'anyi}},
  \bibinfo {author} {\bibfnamefont {S.~R.}\ \bibnamefont {White}}, \bibinfo
  {author} {\bibfnamefont {M.~E.}\ \bibnamefont {Zhitomirsky}},\ and\ \bibinfo
  {author} {\bibfnamefont {A.~L.}\ \bibnamefont {Chernyshev}},\ }\bibfield
  {title} {\bibinfo {title} {{Where is the Quantum Spin Nematic?}},\ }\href
  {https://doi.org/10.1103/PhysRevLett.130.116701} {\bibfield  {journal}
  {\bibinfo  {journal} {Phys. Rev. Lett.}\ }\textbf {\bibinfo {volume} {130}},\
  \bibinfo {pages} {116701} (\bibinfo {year} {2023})}\BibitemShut {NoStop}%
\bibitem [{\citenamefont {Chubukov}(1991)}]{PhysRevB.44.4693}%
  \BibitemOpen
  \bibfield  {author} {\bibinfo {author} {\bibfnamefont {A.~V.}\ \bibnamefont
  {Chubukov}},\ }\bibfield  {title} {\bibinfo {title} {Chiral, nematic, and
  dimer states in quantum spin chains},\ }\href
  {https://doi.org/10.1103/PhysRevB.44.4693} {\bibfield  {journal} {\bibinfo
  {journal} {Phys. Rev. B}\ }\textbf {\bibinfo {volume} {44}},\ \bibinfo
  {pages} {4693} (\bibinfo {year} {1991})}\BibitemShut {NoStop}%
\bibitem [{\citenamefont {Chojnacki}\ \emph {et~al.}(2024)\citenamefont
  {Chojnacki}, \citenamefont {Pohle}, \citenamefont {Yan}, \citenamefont
  {Akagi},\ and\ \citenamefont {Shannon}}]{PhysRevB.109.L220407}%
  \BibitemOpen
  \bibfield  {author} {\bibinfo {author} {\bibfnamefont {L.}~\bibnamefont
  {Chojnacki}}, \bibinfo {author} {\bibfnamefont {R.}~\bibnamefont {Pohle}},
  \bibinfo {author} {\bibfnamefont {H.}~\bibnamefont {Yan}}, \bibinfo {author}
  {\bibfnamefont {Y.}~\bibnamefont {Akagi}},\ and\ \bibinfo {author}
  {\bibfnamefont {N.}~\bibnamefont {Shannon}},\ }\bibfield  {title} {\bibinfo
  {title} {Gravitational wave analogs in spin nematics and cold atoms},\ }\href
  {https://doi.org/10.1103/PhysRevB.109.L220407} {\bibfield  {journal}
  {\bibinfo  {journal} {Phys. Rev. B}\ }\textbf {\bibinfo {volume} {109}},\
  \bibinfo {pages} {L220407} (\bibinfo {year} {2024})}\BibitemShut {NoStop}%
\bibitem [{\citenamefont {Finger}\ \emph {et~al.}(2024)\citenamefont {Finger},
  \citenamefont {Rosa-Medina}, \citenamefont {Reiter}, \citenamefont
  {Christodoulou}, \citenamefont {Donner},\ and\ \citenamefont
  {Esslinger}}]{PhysRevLett.132.093402}%
  \BibitemOpen
  \bibfield  {author} {\bibinfo {author} {\bibfnamefont {F.}~\bibnamefont
  {Finger}}, \bibinfo {author} {\bibfnamefont {R.}~\bibnamefont {Rosa-Medina}},
  \bibinfo {author} {\bibfnamefont {N.}~\bibnamefont {Reiter}}, \bibinfo
  {author} {\bibfnamefont {P.}~\bibnamefont {Christodoulou}}, \bibinfo {author}
  {\bibfnamefont {T.}~\bibnamefont {Donner}},\ and\ \bibinfo {author}
  {\bibfnamefont {T.}~\bibnamefont {Esslinger}},\ }\bibfield  {title} {\bibinfo
  {title} {{Spin- and Momentum-Correlated Atom Pairs Mediated by Photon
  Exchange and Seeded by Vacuum Fluctuations}},\ }\href
  {https://doi.org/10.1103/PhysRevLett.132.093402} {\bibfield  {journal}
  {\bibinfo  {journal} {Phys. Rev. Lett.}\ }\textbf {\bibinfo {volume} {132}},\
  \bibinfo {pages} {093402} (\bibinfo {year} {2024})}\BibitemShut {NoStop}%
\bibitem [{\citenamefont {Cooper}\ \emph {et~al.}(2024)\citenamefont {Cooper},
  \citenamefont {Kunkel}, \citenamefont {Periwal},\ and\ \citenamefont
  {Schleier-Smith}}]{Monika}%
  \BibitemOpen
  \bibfield  {author} {\bibinfo {author} {\bibfnamefont {E.~S.}\ \bibnamefont
  {Cooper}}, \bibinfo {author} {\bibfnamefont {P.}~\bibnamefont {Kunkel}},
  \bibinfo {author} {\bibfnamefont {A.}~\bibnamefont {Periwal}},\ and\ \bibinfo
  {author} {\bibfnamefont {M.}~\bibnamefont {Schleier-Smith}},\ }\bibfield
  {title} {\bibinfo {title} {Graph states of atomic ensembles engineered by
  photon-mediated entanglement},\ }\href
  {https://doi.org/10.1038/s41567-024-02407-1} {\bibfield  {journal} {\bibinfo
  {journal} {Nature Physics}\ }\textbf {\bibinfo {volume} {20}},\ \bibinfo
  {pages} {770} (\bibinfo {year} {2024})}\BibitemShut {NoStop}%
\bibitem [{\citenamefont {Helson}\ \emph {et~al.}(2023)\citenamefont {Helson},
  \citenamefont {Zwettler}, \citenamefont {Mivehvar}, \citenamefont {Colella},
  \citenamefont {Roux}, \citenamefont {Konishi}, \citenamefont {Ritsch},\ and\
  \citenamefont {Brantut}}]{JPhillippe}%
  \BibitemOpen
  \bibfield  {author} {\bibinfo {author} {\bibfnamefont {V.}~\bibnamefont
  {Helson}}, \bibinfo {author} {\bibfnamefont {T.}~\bibnamefont {Zwettler}},
  \bibinfo {author} {\bibfnamefont {F.}~\bibnamefont {Mivehvar}}, \bibinfo
  {author} {\bibfnamefont {E.}~\bibnamefont {Colella}}, \bibinfo {author}
  {\bibfnamefont {K.}~\bibnamefont {Roux}}, \bibinfo {author} {\bibfnamefont
  {H.}~\bibnamefont {Konishi}}, \bibinfo {author} {\bibfnamefont
  {H.}~\bibnamefont {Ritsch}},\ and\ \bibinfo {author} {\bibfnamefont {J.-P.}\
  \bibnamefont {Brantut}},\ }\bibfield  {title} {\bibinfo {title}
  {{Density-wave ordering in a unitary Fermi gas with photon-mediated
  interactions}},\ }\href {https://doi.org/10.1038/s41586-023-06018-3}
  {\bibfield  {journal} {\bibinfo  {journal} {Nature}\ }\textbf {\bibinfo
  {volume} {618}},\ \bibinfo {pages} {716} (\bibinfo {year}
  {2023})}\BibitemShut {NoStop}%
\bibitem [{\citenamefont {Chen}\ and\ \citenamefont
  {Frisk~Kockrum}(2026)}]{Frisk}%
  \BibitemOpen
  \bibfield  {author} {\bibinfo {author} {\bibfnamefont {G.}~\bibnamefont
  {Chen}}\ and\ \bibinfo {author} {\bibfnamefont {A.}~\bibnamefont
  {Frisk~Kockrum}},\ }\bibfield  {title} {\bibinfo {title} {Scalable quantum
  simulator with an extended gate set in giant atoms},\ }\href
  {https://doi.org/10.22331/q-2026-01-30-1992} {\bibfield  {journal} {\bibinfo
  {journal} {Quantum}\ }\textbf {\bibinfo {volume} {10}},\ \bibinfo {pages}
  {1992} (\bibinfo {year} {2026})}\BibitemShut {NoStop}%
\bibitem [{\citenamefont {Kozlowski}\ \emph {et~al.}(2015)\citenamefont
  {Kozlowski}, \citenamefont {Caballero-Benitez},\ and\ \citenamefont
  {Mekhov}}]{PhysRevA.92.013613}%
  \BibitemOpen
  \bibfield  {author} {\bibinfo {author} {\bibfnamefont {W.}~\bibnamefont
  {Kozlowski}}, \bibinfo {author} {\bibfnamefont {S.~F.}\ \bibnamefont
  {Caballero-Benitez}},\ and\ \bibinfo {author} {\bibfnamefont {I.~B.}\
  \bibnamefont {Mekhov}},\ }\bibfield  {title} {\bibinfo {title} {Probing
  matter-field and atom-number correlations in optical lattices by global
  nondestructive addressing},\ }\href
  {https://doi.org/10.1103/PhysRevA.92.013613} {\bibfield  {journal} {\bibinfo
  {journal} {Phys. Rev. A}\ }\textbf {\bibinfo {volume} {92}},\ \bibinfo
  {pages} {013613} (\bibinfo {year} {2015})}\BibitemShut {NoStop}%
\bibitem [{\citenamefont {White}(1992)}]{DMRG1}%
  \BibitemOpen
  \bibfield  {author} {\bibinfo {author} {\bibfnamefont {S.~R.}\ \bibnamefont
  {White}},\ }\bibfield  {title} {\bibinfo {title} {Density matrix formulation
  for quantum renormalization groups},\ }\href
  {https://doi.org/10.1103/PhysRevLett.69.2863} {\bibfield  {journal} {\bibinfo
   {journal} {Phys. Rev. Lett.}\ }\textbf {\bibinfo {volume} {69}},\ \bibinfo
  {pages} {2863} (\bibinfo {year} {1992})}\BibitemShut {NoStop}%
\bibitem [{\citenamefont {Schollw\"ock}(2005)}]{RevModPhys.77.259}%
  \BibitemOpen
  \bibfield  {author} {\bibinfo {author} {\bibfnamefont {U.}~\bibnamefont
  {Schollw\"ock}},\ }\bibfield  {title} {\bibinfo {title} {The density-matrix
  renormalization group},\ }\href {https://doi.org/10.1103/RevModPhys.77.259}
  {\bibfield  {journal} {\bibinfo  {journal} {Rev. Mod. Phys.}\ }\textbf
  {\bibinfo {volume} {77}},\ \bibinfo {pages} {259} (\bibinfo {year}
  {2005})}\BibitemShut {NoStop}%
\bibitem [{\citenamefont {Fishman}\ \emph {et~al.}(2022)\citenamefont
  {Fishman}, \citenamefont {White},\ and\ \citenamefont
  {Stoudenmire}}]{itensor2}%
  \BibitemOpen
  \bibfield  {author} {\bibinfo {author} {\bibfnamefont {M.}~\bibnamefont
  {Fishman}}, \bibinfo {author} {\bibfnamefont {S.~R.}\ \bibnamefont {White}},\
  and\ \bibinfo {author} {\bibfnamefont {E.~M.}\ \bibnamefont {Stoudenmire}},\
  }\bibfield  {title} {\bibinfo {title} {{Codebase release 0.3 for ITensor}},\
  }\href {https://doi.org/10.21468/SciPostPhysCodeb.4-r0.3} {\bibfield
  {journal} {\bibinfo  {journal} {SciPost Phys. Codebases}\ ,\ \bibinfo {pages}
  {4}} (\bibinfo {year} {2022})}\BibitemShut {NoStop}%
\bibitem [{\citenamefont {Giamarchi}()}]{Giamarchi}%
  \BibitemOpen
  \bibfield  {author} {\bibinfo {author} {\bibfnamefont {T.}~\bibnamefont
  {Giamarchi}},\ }\href@noop {} {\emph {\bibinfo {title} {{Quantum Physics in
  One Dimension},}}}\BibitemShut {NoStop}%
\bibitem [{\citenamefont {Haldane}(1981)}]{Haldane}%
  \BibitemOpen
  \bibfield  {author} {\bibinfo {author} {\bibfnamefont {F.~D.~M.}\
  \bibnamefont {Haldane}},\ }\bibfield  {title} {\bibinfo {title} {{Luttinger
  liquid theory of one-dimensional quantum fluids. I. Properties of the
  Luttinger model and their extension to the general 1D interacting Fermi
  gas}},\ }\href {https://doi.org/10.1088/0022-3719/14/19/010} {\bibfield
  {journal} {\bibinfo  {journal} {J. Phys. C: Solid State Phys.}\ }\textbf
  {\bibinfo {volume} {14}},\ \bibinfo {pages} {2585} (\bibinfo {year}
  {1981})}\BibitemShut {NoStop}%
\bibitem [{\citenamefont {Camacho-Guardian}\ \emph {et~al.}(2017)\citenamefont
  {Camacho-Guardian}, \citenamefont {Paredes},\ and\ \citenamefont
  {Caballero-Ben\'{\i}tez}}]{PhysRevA.96.051602}%
  \BibitemOpen
  \bibfield  {author} {\bibinfo {author} {\bibfnamefont {A.}~\bibnamefont
  {Camacho-Guardian}}, \bibinfo {author} {\bibfnamefont {R.}~\bibnamefont
  {Paredes}},\ and\ \bibinfo {author} {\bibfnamefont {S.~F.}\ \bibnamefont
  {Caballero-Ben\'{\i}tez}},\ }\bibfield  {title} {\bibinfo {title} {Quantum
  simulation of competing orders with fermions in quantum optical lattices},\
  }\href {https://doi.org/10.1103/PhysRevA.96.051602} {\bibfield  {journal}
  {\bibinfo  {journal} {Phys. Rev. A}\ }\textbf {\bibinfo {volume} {96}},\
  \bibinfo {pages} {051602(R)} (\bibinfo {year} {2017})}\BibitemShut {NoStop}%
\bibitem [{\citenamefont {Mazzucchi}\ \emph
  {et~al.}(2016{\natexlab{b}})\citenamefont {Mazzucchi}, \citenamefont
  {Caballero-Benitez},\ and\ \citenamefont {Mekhov}}]{Mazzucchi2016}%
  \BibitemOpen
  \bibfield  {author} {\bibinfo {author} {\bibfnamefont {G.}~\bibnamefont
  {Mazzucchi}}, \bibinfo {author} {\bibfnamefont {S.~F.}\ \bibnamefont
  {Caballero-Benitez}},\ and\ \bibinfo {author} {\bibfnamefont {I.~B.}\
  \bibnamefont {Mekhov}},\ }\bibfield  {title} {\bibinfo {title} {{Quantum
  measurement-induced antiferromagnetic order and density modulations in
  ultracold Fermi gases in optical lattices}},\ }\href
  {https://doi.org/10.1038/srep31196} {\bibfield  {journal} {\bibinfo
  {journal} {Scientific Reports}\ }\textbf {\bibinfo {volume} {6}},\ \bibinfo
  {pages} {31196} (\bibinfo {year} {2016}{\natexlab{b}})}\BibitemShut {NoStop}%
\bibitem [{\citenamefont {Caballero-Benitez}(2026)}]{MajoranaQOL}%
  \BibitemOpen
  \bibfield  {author} {\bibinfo {author} {\bibfnamefont {S.~F.}\ \bibnamefont
  {Caballero-Benitez}},\ }\href {https://arxiv.org/abs/2408.08559} {\bibinfo
  {title} {{Stability of Majorana zero modes with quantum optical lattices}}}
  (\bibinfo {year} {2026}),\ \Eprint {https://arxiv.org/abs/2408.08559}
  {arXiv:2408.08559 [cond-mat.quant-gas]} \BibitemShut {NoStop}%
\bibitem [{\citenamefont {Lewenstein}\ \emph {et~al.}(2012)\citenamefont
  {Lewenstein}, \citenamefont {Sanpera},\ and\ \citenamefont
  {Ahufinger}}]{Lewenstein2012}%
  \BibitemOpen
  \bibfield  {author} {\bibinfo {author} {\bibfnamefont {M.}~\bibnamefont
  {Lewenstein}}, \bibinfo {author} {\bibfnamefont {A.}~\bibnamefont
  {Sanpera}},\ and\ \bibinfo {author} {\bibfnamefont {V.}~\bibnamefont
  {Ahufinger}},\ }\href@noop {} {\emph {\bibinfo {title} {{Ultracold Atoms in
  Optical Lattices: Simulating Quantum Many-Body Systems}}}}\ (\bibinfo
  {publisher} {Oxford University Press},\ \bibinfo {address} {Oxford},\
  \bibinfo {year} {2012})\BibitemShut {NoStop}%
\bibitem [{\citenamefont {Biamonte}\ and\ \citenamefont
  {Love}(2008)}]{Biamonte2008}%
  \BibitemOpen
  \bibfield  {author} {\bibinfo {author} {\bibfnamefont {S.~S.}\ \bibnamefont
  {Biamonte}}\ and\ \bibinfo {author} {\bibfnamefont {P.~J.}\ \bibnamefont
  {Love}},\ }\bibfield  {title} {\bibinfo {title} {Realizing nearest-neighbor
  quantum circuits on one-dimensional architectures},\ }\href
  {https://doi.org/10.1103/PhysRevA.78.012352} {\bibfield  {journal} {\bibinfo
  {journal} {Physical Review A}\ }\textbf {\bibinfo {volume} {78}},\ \bibinfo
  {pages} {012352} (\bibinfo {year} {2008})}\BibitemShut {NoStop}%
\bibitem [{\citenamefont {Li}\ \emph {et~al.}(2019)\citenamefont {Li},
  \citenamefont {Ding},\ and\ \citenamefont {Xie}}]{Li2019}%
  \BibitemOpen
  \bibfield  {author} {\bibinfo {author} {\bibfnamefont {G.}~\bibnamefont
  {Li}}, \bibinfo {author} {\bibfnamefont {Y.}~\bibnamefont {Ding}},\ and\
  \bibinfo {author} {\bibfnamefont {Y.}~\bibnamefont {Xie}},\ }\bibfield
  {title} {\bibinfo {title} {{Tackling the qubit mapping problem for NISQ-era
  quantum devices}},\ }in\ \href {https://doi.org/10.1145/3297858.3304023}
  {\emph {\bibinfo {booktitle} {Proceedings of the Twenty-Fourth International
  Conference on Architectural Support for Programming Languages and Operating
  Systems (ASPLOS)}}}\ (\bibinfo {organization} {ACM},\ \bibinfo {year}
  {2019})\ pp.\ \bibinfo {pages} {1001--1014}\BibitemShut {NoStop}%
\bibitem [{\citenamefont {Briegel}\ \emph {et~al.}(2009)\citenamefont
  {Briegel}, \citenamefont {Browne}, \citenamefont {D{\"u}r}, \citenamefont
  {Raussendorf},\ and\ \citenamefont {Van~den Nest}}]{Briegel2009MBQC}%
  \BibitemOpen
  \bibfield  {author} {\bibinfo {author} {\bibfnamefont {H.~J.}\ \bibnamefont
  {Briegel}}, \bibinfo {author} {\bibfnamefont {D.~E.}\ \bibnamefont {Browne}},
  \bibinfo {author} {\bibfnamefont {W.}~\bibnamefont {D{\"u}r}}, \bibinfo
  {author} {\bibfnamefont {R.}~\bibnamefont {Raussendorf}},\ and\ \bibinfo
  {author} {\bibfnamefont {M.}~\bibnamefont {Van~den Nest}},\ }\bibfield
  {title} {\bibinfo {title} {Measurement-based quantum computation},\ }\href
  {https://doi.org/10.1038/nphys1168} {\bibfield  {journal} {\bibinfo
  {journal} {Nature Physics}\ }\textbf {\bibinfo {volume} {5}},\ \bibinfo
  {pages} {19} (\bibinfo {year} {2009})}\BibitemShut {NoStop}%
\bibitem [{\citenamefont {Saffman}(2016)}]{Saffman2016NeutralAtoms}%
  \BibitemOpen
  \bibfield  {author} {\bibinfo {author} {\bibfnamefont {M.}~\bibnamefont
  {Saffman}},\ }\bibfield  {title} {\bibinfo {title} {Quantum computing with
  neutral atoms},\ }\href {https://doi.org/10.1088/0953-4075/49/20/202001}
  {\bibfield  {journal} {\bibinfo  {journal} {Journal of Physics B: Atomic,
  Molecular and Optical Physics}\ }\textbf {\bibinfo {volume} {49}},\ \bibinfo
  {pages} {202001} (\bibinfo {year} {2016})}\BibitemShut {NoStop}%
\bibitem [{\citenamefont {Tannu}\ and\ \citenamefont
  {Qureshi}(2019)}]{Tannu2019Compilation}%
  \BibitemOpen
  \bibfield  {author} {\bibinfo {author} {\bibfnamefont {S.~S.}\ \bibnamefont
  {Tannu}}\ and\ \bibinfo {author} {\bibfnamefont {M.~K.}\ \bibnamefont
  {Qureshi}},\ }\bibfield  {title} {\bibinfo {title} {{Quantum circuit
  compilation and optimization for NISQ architectures}},\ }\href
  {https://doi.org/10.1109/MM.2019.2942979} {\bibfield  {journal} {\bibinfo
  {journal} {IEEE Micro}\ }\textbf {\bibinfo {volume} {39}},\ \bibinfo {pages}
  {72} (\bibinfo {year} {2019})}\BibitemShut {NoStop}%
\end{thebibliography}%

\pagebreak
\clearpage
\newpage
\widetext
\begin{center}
\textbf{\large Supplemental Material for: Quantum Correlations and Entanglement in Generalized Dicke-Ising Models}
\\
{Santiago F. Caballero-Benitez$^{1}$}%
 \\
{%
$^{1}$Instituto de Física, LSCSC-LANMAC, Universidad Nacional Autónoma de México, Ciudad de México 04510, Mexico
}%
\end{center}
\setcounter{equation}{0}
\setcounter{figure}{0}
\setcounter{table}{0}
\setcounter{page}{1}
\makeatletter
\renewcommand{\theequation}{S\arabic{equation}}
\renewcommand{\thefigure}{S\arabic{figure}}

\section{Cavity back action and light matter overlap strengths}
Due to the different time scales of the spin and the cavity field dynamics, one can adiabatically eliminate the latter similar to atomic systems\cite{doi:10.1080/00018732.2021.1969727}. If the energy scales are not that separated the scheme can be amended by considering additional contributions as in~\cite{NewJPhys.17.123023}.
The photonic operator $\hat a$ is replaced by its  steady state value $\alpha=\braket{\hat a}$. This is found  taking the steady state value from the Heisenberg equation of motion for the light field: $d\braket{\hat{a}}/dt = (i/\hbar)\braket{[\mathcal{H},\hat{a}]}- \kappa\braket{\hat{a}}=0$.   Here the dissipation from the cavity is introduced phenomenologically through the cavity decay rate $\kappa$. It follows that,
\begin{equation}\label{eq:alpha}
    \alpha =\frac{\hbar V_p}{(\hbar\Delta_c +i\hbar \kappa)\sqrt{N}}\sum_nJ^{pc}_n\langle\hat{S}^x_n\rangle
\end{equation}
which is solved self consistently as function of $V_p$,  $\Delta_c$. and $\kappa$. Note that in the SR state the phase of the light field is $\pi$ in this setup.
Following~\cite{PhysRevA.92.013613}, the overlap integrals $J^{pc}_i$ can be written in terms of the incident pump field,  the overlap with the cavity mode and the localized spins at position $n$ as,
\begin{eqnarray}
J^{pc}_n&=&\int d x |w(x-x_n)|^2\cos(\textbf{k}_p\cdot\textbf{x})\cos(\textbf{k}_c\cdot\textbf{x})
\\
&\approx&\cos(\pi n)\cos(\pi n\cos\phi)
\end{eqnarray} 
with $w(x)$ the effective Wannier function that describes the localization of the each spin at position $x$, with $\phi$ the angle between the cavity axis and the pump wave-vector. Note that the pump, the cavity mode and the spins positions have been chosen commensurate to each other fixing $|\textbf{k}_p|=|\textbf{k}_c|=2\pi/\lambda$ with the lattice spacing between spins as $a=\lambda/2$ and $x_n=n a$.

{\section{Self Consistent Algorithm: light-matter DMRG for spins}\label{sec:SelfConsistentAlgorithm}} 

Similar to the light-matter DMRG algorithm for ultracold atoms~\cite{gvm2-b46t} we employ the same structure for the spin system, a key difference here is that the Wannier functions have been approximated static (not dependent on the cavity light) without loss of generality, further refinements for higher dimensions can be implemented as well as the full dependence on cavity light.
The phase diagrams are obtained solving \eqref{eq:alpha} self consistently. We start giving an ansatz $\alpha\neq 0$, we obtain the ground state of $\mathcal{H}$ substituting  $\hat a\to\alpha$ and $\hat a^\dagger\to\alpha^*$. Then we obtain a new value for $\alpha$\ from \eqref{eq:alpha}. This cycle is repeated until convergence is achieved with a tolerance less than single precision. There can be multiple solutions for $\alpha$ for fixed $V_p$,  $\Delta_c$ and $\kappa$, so it is necessary to keep only the lowest energy  state. 

 To find the effective ground state of (\ref{model}) for a fixed parameter set in the self-consistent loop, we perform simulations using DMRG~\cite{DMRG1,RevModPhys.77.259} with the aid of the ITensor library\cite{ itensor2}. We simulate 1D chains with up to 1025 sites with open boundary conditions; we verified lower number of sites recover qualitative similar results.

\begin{figure}[t!]
    \centering
    \includegraphics[width=0.47\textwidth]{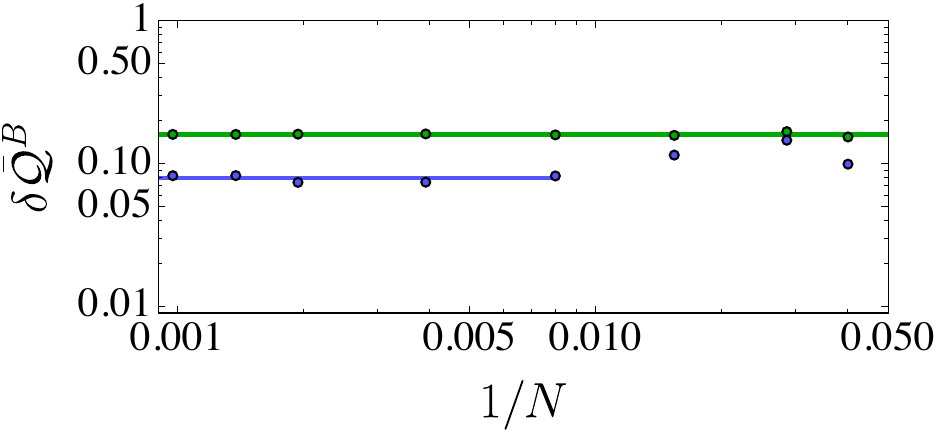}

    \caption{
     \textbf{Finite size analysis of the quantum correlations of nematic order $\delta\bar{\mathcal{Q}}^B$ in the $\varphi$ mode induced by light near criticality}. As the number of sites increases  $\delta\bar{\mathcal{Q}}^B$ becomes constant for both $J>1$ (blue) and $J<1$ (green), signalling  true long range bond nematic order. The solid lines are the constant fit limit value of $\delta\bar{\mathcal{Q}}^B=$ 0.078 ($J>1$), 0.160 ($J<1$), the extent of the lines is the data points considered. The parameters are: $n=N/2$ the middle of the chain, $\omega_0/|J|=0.1$, $\kappa/|J|=10$, $\Delta_c/|J|=-50$ and $V_p$ at the maximum possible $\tilde{\mathcal{Q}}^B$ near the transition to the SR phase. The number of sites spans $N\in[1025,25]$. 
     }
    \label{fig:Fig4a}
\end{figure}

\begin{figure}[t!]
    \centering
    \includegraphics[width=0.47\textwidth]{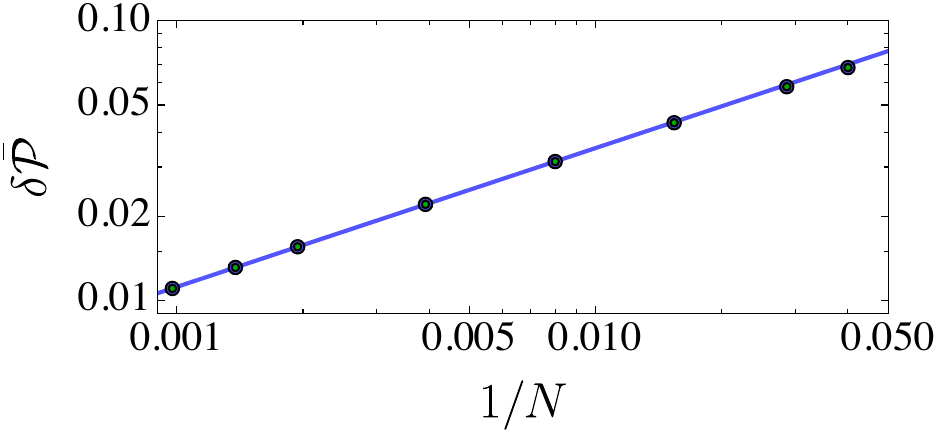}

    \caption{
 \textbf{Finite size analysis of the quantum correlations of magnon pair amplitude $\delta\bar{\mathcal{P}}$ in the $\varphi$ mode induced by light near criticality}. As the number of sites increases  $\delta\bar{\mathcal{P}}$ exhibit quasi-long range order $J>1$ (blue) and $J<1$ (green), signalling  true long range bond nematic order. The solid lines are the fit $\delta\bar{\mathcal{P}}= 0.34N^{-0.497}$  ($J>0$), $0.33N^{-0.490}$ ($J<0$), exhibiting the characteristic behaviour of Luttinger theory. The parameters are the same as Fig.\ref{fig:Fig4a}.
     }
    \label{fig:Fig4c}
\end{figure}

\begin{figure}[t!]
    \centering
    \includegraphics[width=0.47\textwidth]{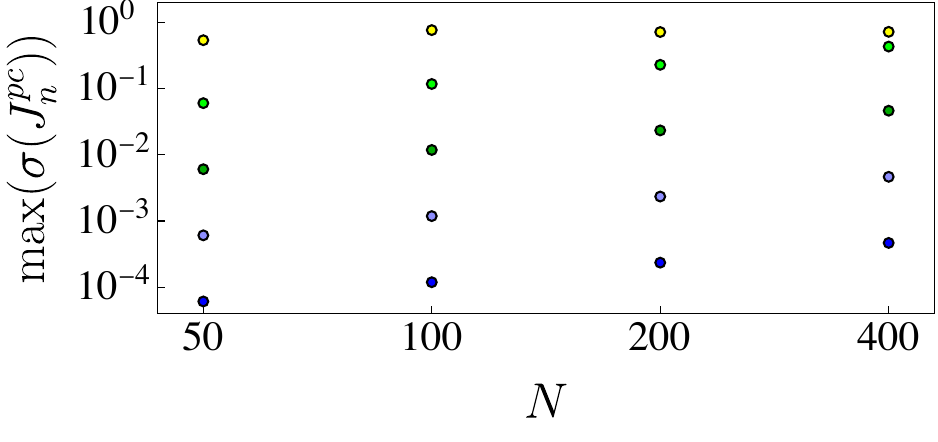}

    \caption{
    \textcolor{black}{ \textbf{Sensitivity of the angle to generate the golden ratio modes $\varphi$}. 
     The maximum standard deviation of the set of $J^{pc}_n$ couplings for $n\in\{1,\dots, 5\}$ for different lattice sizes $N$. We consider variations in the couplings that span the values from top to bottom $\{10^{-2},10^{-3},10^{-4},10^{-5},10^{-6}\}$.
     }}
    \label{fig:Fig4b}
\end{figure}

 {\section{Finite size scaling of the quantum correlations of the bond nematic order $\mathcal{Q}^B$  and the pairing amplitude $\mathcal{P}$}} 

  We perform finite size analysis near the critical point of the SR transition for the $\varphi$ setting that maximizes bond nematic order. We calculate (the connected correlation)
 \begin{equation}
 \delta\bar{\mathcal{O}}=\sqrt{\frac{2}{N}\sum_{l=-N/4}^{N/4}(\delta\mathcal{O}_{\frac{N}{2} \frac{N}{2}+l})^2}, 
 \end{equation}
 computing the bulk value and we have eliminated boundary effects, as we are using open boundary conditions and $\mathcal{O}\in\{\mathcal{Q}^B ,\mathcal{P}\}$. The key signal in this analysis is the fact that $\delta\bar{\mathcal{Q}}^B$ is independent of system size, there is no algebraic decay, Fig.\ref{fig:Fig4a}. On the other hand,  $\delta\bar{\mathcal{P}}\sim {N}^{-1/2+\epsilon}$ exhibits algebraic decay proper to quasi-long range order, with $\epsilon\neq0$ originating from the periodic structure induced by light,  Fig.\ref{fig:Fig4c}. The Luttinger parameters~\cite{Giamarchi,Haldane} are $K=0.993$ ($J>0$), $0.982$ ($J<0$). Therefore, we have true long range bond nematic order that can coexist with dipolar order if $\mathcal{M}^x_\theta\neq 0$ simultaneously.

 {\section{Sensitivity of the angle variation for multimode light-induced configurations ($\varphi$)}}  
 
  The stability required in the control of the variation $\delta\phi$ of the angle  $\phi$ of the light pumped into the system is estimated by considering the largest deviation in the set of the coupling values $J_{n}^{pc}$ in the the five site set of the golden ratio scheme, depending on the number of sites $N$. Thus, the sensitivity $S(\delta\phi)\sim\mathrm{max}(\sigma(J^{pc}_n))$ for each set of five $\varphi$ couplings, with $\sigma(\cdot)$ the standard deviation across a lattice of $N$ sites, see Fig.\ref{fig:Fig4b}. We take the set of sites $n\in[-N/2,N/2-1]$,  $|\delta\phi|\in[10^{-2},10^{-6}]$ and $N\in\{50,100,200,400\}$. An immediate consequence of this is that the precision required to generate accurately larger commensurate multimode configurations is limited by the system size.
\\

\section{Details for realizations of the scheme in different physical systems.} 
 Typical models that describe Rydberg atoms inside a cavity are similar to our setting, see \cite{SciPostPhys.1.1.004,CavityRydberg,Rydberg1}. Considering the atoms with additional contributions 
beyond nearest neighbors the interaction would become,
\begin{equation}
\mathcal{H}^{\mathrm{Ryb}}_{\mathrm{Int}}=\sum_{\langle n,m\rangle}g_{n2}\hat{S}^z_n\hat{S}^z_m+\sum_{\langle\langle n,m\rangle\rangle}g_{n3}\hat{S}^z_n\hat{S}^z_m+\dots 
\end{equation}
where $g_{n2}$ is the coupling between nearest neighbors in the lattice\cite{SciPostPhys.1.1.004} and $g_{n3}$  is the coupling to next nearest neighbors. The couplings $g_{n2|3}$ are the effective dipole-dipole interaction that depends on the density and separation between the atoms depending on the experimental setting, typically $g_{n2}\gg g_{n3}$ and subsequent more separated neighbors, all the other terms in the model are the same. Cold fermions inside a cavity would entail the combination of the scheme in \cite{PhysRevResearch.3.013178} and \cite{Rey}, where we would have the Hamiltonian for the cold fermions with spin-orbit (SOC) coupling:
\begin{equation}
\mathcal{H}^{\mathrm{CF}}_{\mathrm{eff}} = \sum_{\langle n,m\rangle}\left( J_x\hat{S}_n^x \hat{S}_{m}^x + J_y\hat{S}_n^y \hat{S}_{m}^y \right) + J_{\perp} \sum_{\langle n,m\rangle} \hat{S}_n^z \hat{S}_{m}^z 
 +  J_{\Omega} \sum_{n} \hat{S}_n^x
\end{equation}
and possibly it would be necessary to include spontaneous photon emission with rate $\gamma$ via the dissipator: 
\begin{equation}
\mathcal{L}_{\textrm{CF}}(\hat{\rho}) = \frac{\gamma}{2} \sum_{n}\left( 2 \hat{S}_n^- \hat{\rho} \hat{S}_n^+ - \hat{S}_n^+ \hat{S}_n^- \hat{\rho} - \hat{\rho} \hat{S}_n^+ \hat{S}_n^- \right),
\end{equation} 
with $\hat\rho$ the density matrix in addition to our light-matter scheme.
Considering ultracold fermions, can be done similar to\cite{PhysRevA.96.051602, Mazzucchi2016, MajoranaQOL}. Specifically, considering~\cite{MajoranaQOL} and performing Jordan-Wigner transformation~\cite{Lewenstein2012,Sachdev2011} to spins, it follows:
\begin{equation}
\mathcal{H} ^{\mathrm{F}}_{\mathrm{eff}}=-2t \sum_{n} \left( \hat{S}_n^x \hat{S}_{n+1}^x + \hat{S}_n^y \hat{S}_{n+1}^y \right) - g_p \sum_{n} \hat{S}_n^z \hat{S}_{n+1}^z
 - g_p \sum_{n} \hat{S}_n^z,
\end{equation}
with $t$ the tunneling amplitude of the fermions in the lattice and $g_p$ the effective $p-$wave interaction. Now the light-matter interacting term would change to,
\begin{equation}
\mathcal{H} ^{\mathrm{F}}_{\mathrm{lm}}=\frac{\hbar V_p}{\sqrt{N}}\left(\hat{a}^\dagger+\hat{a}\right)\sum_nJ^{pc}_n\hat{S}^z_n.
\end{equation}

\section{Order parameters for different setups}
\begin{figure}[bh!]
    \centering
   \includegraphics[scale=0.65]{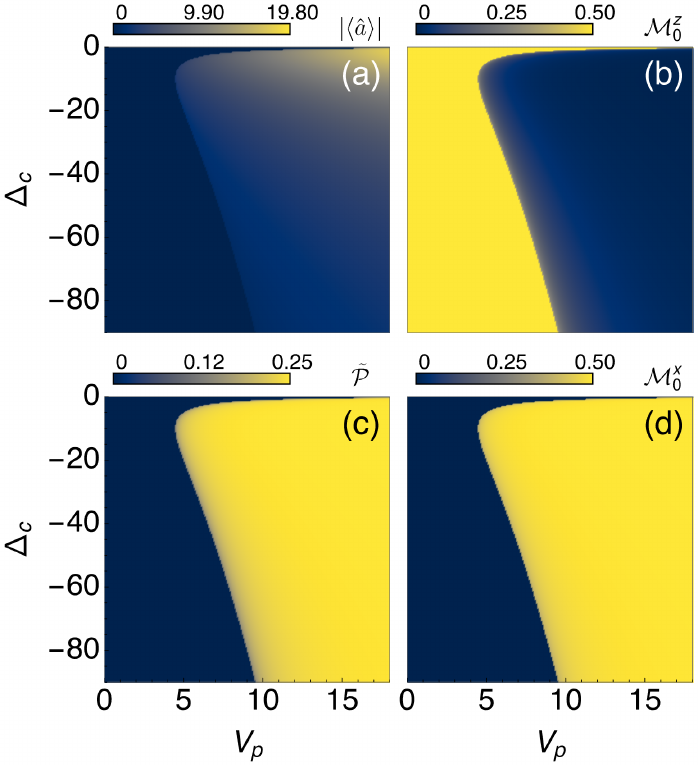}
    \caption{\textbf{Order parameters for normal Ferromagnetic ordering and Superradiant Ferromagnetic setup.}
    (\textbf{a}) Cavity light,  (\textbf{b}) magnetization along quantization axis $``z"$, (\textbf{c}) mean magnon pairing amplitude, (\textbf{d}) magnetization along quantization axis $``x"$. The Ising coupling is $J<0$ FM.
     }
    \label{fig:supp1}
\end{figure}

\begin{figure}[th!]
    \centering
   \includegraphics[scale=0.65]{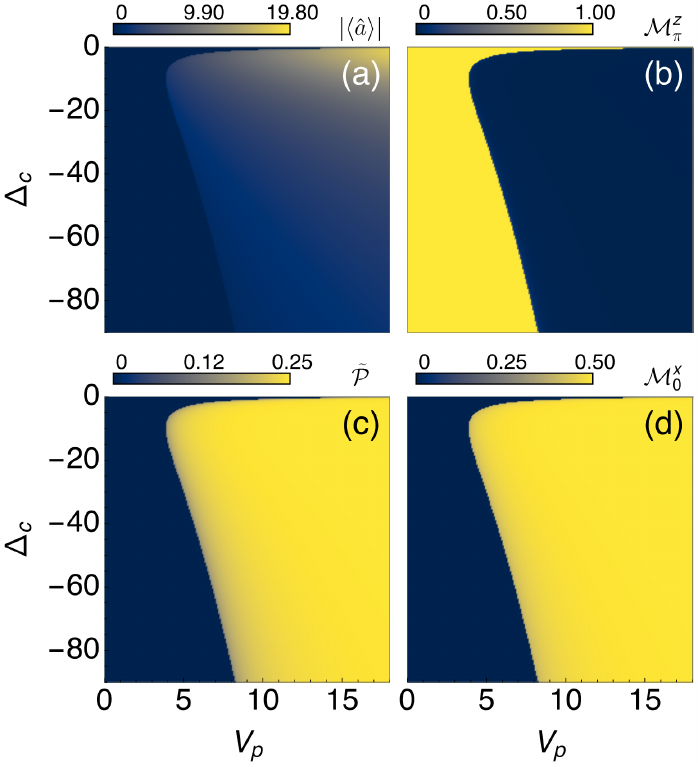}
    \caption{\textbf{Order parameters for normal Anti-ferromagnetic ordering and Superradiant Ferromagnetic setup.}
    \ (\textbf{a}) Cavity light,  (\textbf{b}) staggered magnetization along quantization axis $``z"$, (\textbf{c}) mean magnon pairing amplitude, (\textbf{d}) magnetization along quantization axis $``x"$. The Ising coupling is $J>0$ AFM.
     }
    \label{fig:supp2}
\end{figure}

\begin{figure}[bh!]
    \centering
   \includegraphics[scale=0.65]{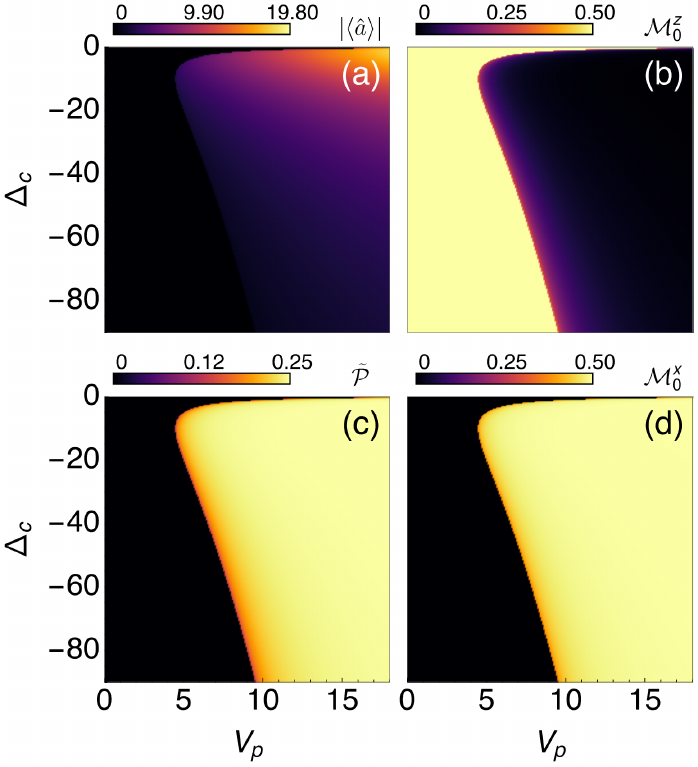}
    \caption{\textbf{Order parameters for normal Ferromagnetic ordering and Superradiant Anti-ferromagnetic setup.}
  \ (\textbf{a}) Cavity light,  (\textbf{b}) magnetization along quantization axis $``z"$, (\textbf{c}) mean magnon pairing amplitude, (\textbf{d}) magnetization along quantization axis $``x"$. The Ising coupling is $J>0$ AFM.
     }
    \label{fig:supp3}
\end{figure}

\begin{figure}[th!]
    \centering
   \includegraphics[scale=0.65]{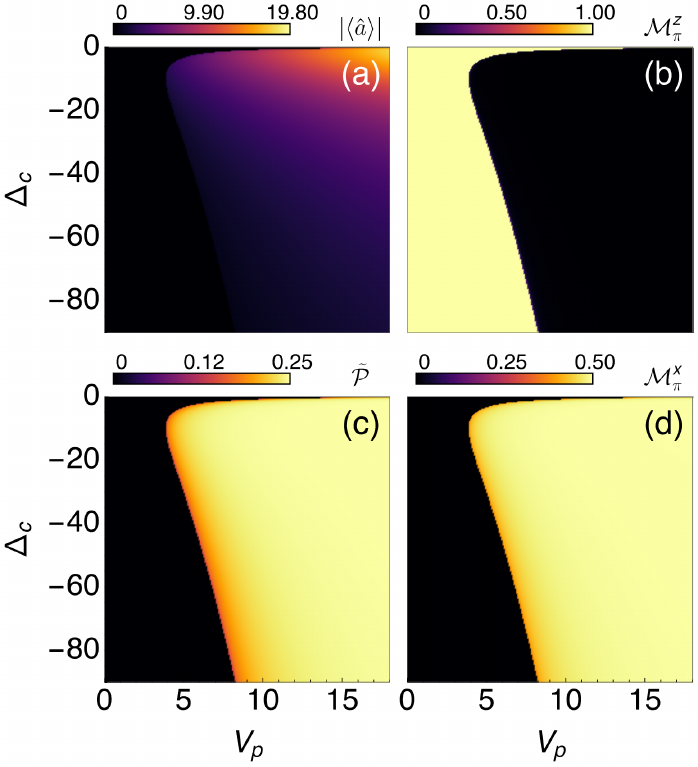}
    \caption{\textbf{Order parameters for normal Anti-ferromagnetic ordering and Superradiant Anti-ferromagnetic setup.}
      \ (\textbf{a}) Cavity light,  (\textbf{b}) staggered magnetization along quantization axis $``z"$, (\textbf{c}) mean magnon pairing amplitude, (\textbf{d}) staggered magnetization along quantization axis $``x"$. The Ising coupling is $J>0$ AFM.
     }
    \label{fig:supp4}
\end{figure}

\begin{figure}[bh!]
    \centering
   \includegraphics[scale=0.65]{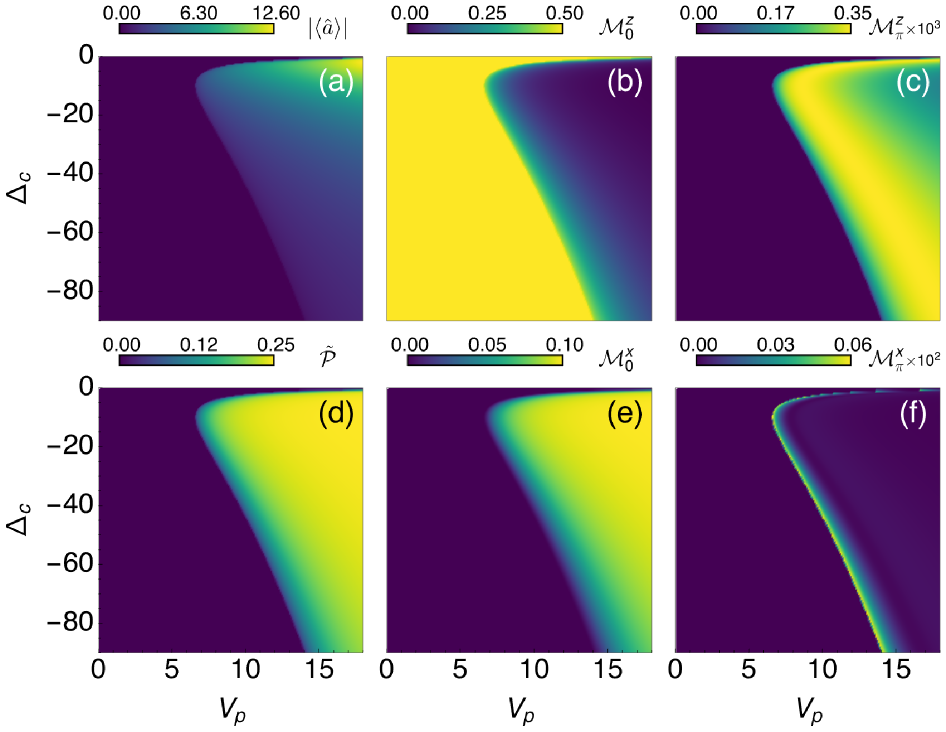}
    \caption{\textbf{Order parameters for normal Ferromagnetic ordering and Superradiant Bond-nematic setup.}
      \  (\textbf{a}) Cavity light,  (\textbf{b}) magnetization along quantization axis $``z"$,  (\textbf{c}) staggered  magnetization along quantization axis $``z"$(\textbf{d}) mean magnon pairing amplitude, (\textbf{e})  magnetization along quantization axis $``x"$, (\textbf{e})  staggered magnetization along quantization axis $``x"$. The Ising coupling is $J<0$ FM.  
        }
    \label{fig:supp5}
\end{figure}

\begin{figure}[th!]
    \centering
   \includegraphics[scale=0.65]{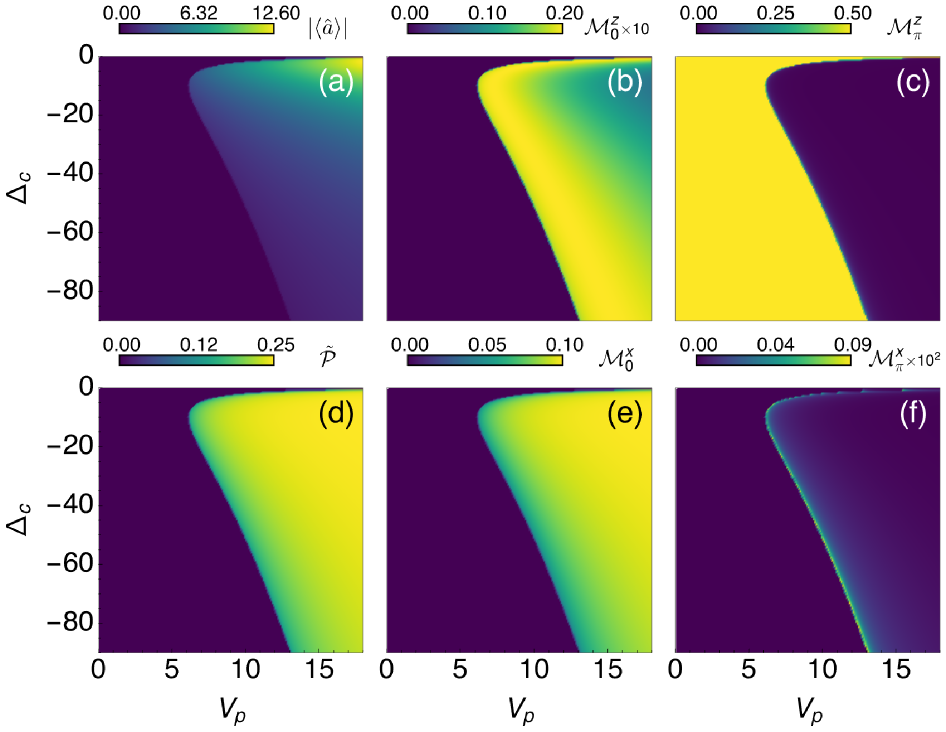}
    \caption{\textbf{Order parameters for normal Anti-ferromagnetic ordering and Superradiant Bond-nematic setup.}
          \  (\textbf{a}) Cavity light,  (\textbf{b}) magnetization along quantization axis $``z"$,  (\textbf{c}) staggered  magnetization along quantization axis $``z"$(\textbf{d}) mean magnon pairing amplitude, (\textbf{e})  magnetization along quantization axis $``x"$, (\textbf{e})  staggered magnetization along quantization axis $``x"$. The Ising coupling is $J>0$ AFM.  
       }
    \label{fig:supp6}
\end{figure}

\begin{figure}[bh!]
    \centering
   \includegraphics[scale=0.65]{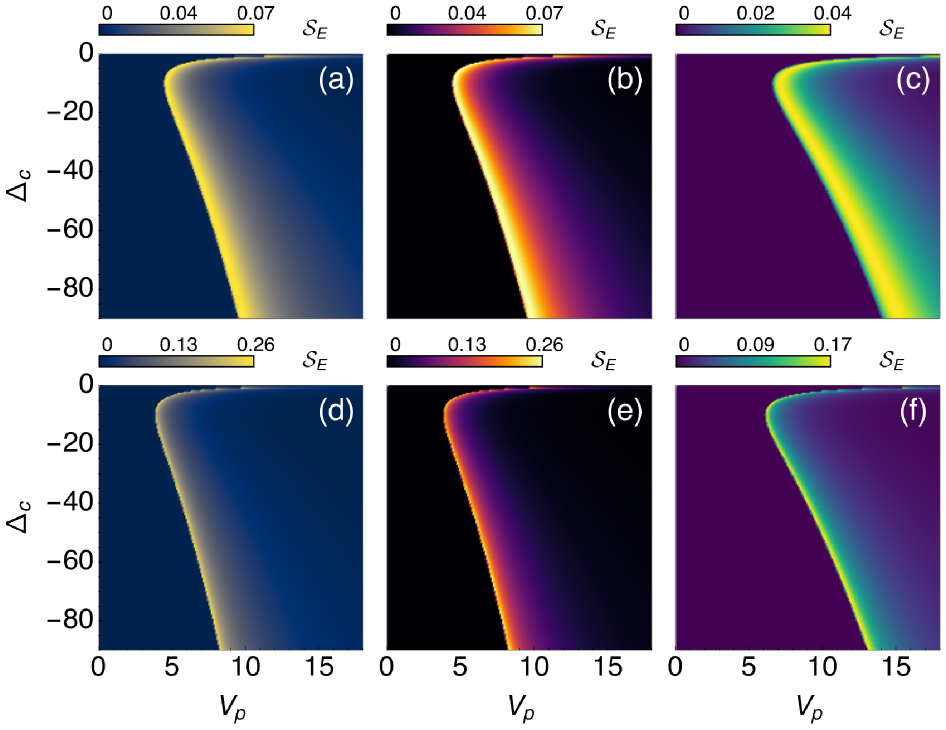}
    \caption{\textbf{Entanglement entropy across half the chain for different configurations.}
      \ The Entanglement entropy corresponds to the setups in Fig. \ref{fig:Fig2}. As the SR transition is approached entanglement is maximal.
      Parameters are: (\textbf{a}) $J<0$, $\phi=0$; (\textbf{b}) $J<0$, $\phi=\pi/2$; (\textbf{c}) $J<0$, $\phi=\arccos(1/5)$; (\textbf{d}) $J>0$,$ \phi=0$; (\textbf{e}) $J>0$, $\phi=\pi/2$; (\textbf{f}) $J>0$, $\phi=\arccos(1/5)$;
     }
    \label{fig:supp7}
\end{figure}

In Fig. \ref{fig:supp1} we show the behavior of the order parameters for $J<1$ and $\phi=0$.
In Fig. \ref{fig:supp2} we show the behavior of the order parameters for $J>1$ and $\phi=0$.
In Fig. \ref{fig:supp3} we show the behavior of the order parameters for $J<1$ and $\phi=\pi/2$.
In Fig. \ref{fig:supp4} we show the behavior of the order parameters for $J>1$ and $\phi=\pi/2$.
In Fig. \ref{fig:supp5} we show the behavior of the order parameters for $J<1$ and $\phi=\arccos(1/5)$.
In Fig. \ref{fig:supp6} we show the behavior of the order parameters for $J>1$ and $\phi=\arccos(1/5)$.
In Fig. \ref{fig:supp7} we show the behavior of the entanglement across the half the chain corresponding to Fig. \ref{fig:Fig2}.
The parameters in all plots are $N=400$ sites with open boundary conditions using light-matter DMRG, $\omega/|J|=0.1$,  $\kappa/|J|=10$, with $\Delta_c$ and $V_p$ in units of $|J|$. Only the non-trivial order parameters are displayed for each setup.

\clearpage
\newpage
\textcolor{black}{
\section{Equivalent circuit model for entanglement generation with the $\varphi$ scheme}
Following the methods in~\cite{Biamonte2008, Li2019} we construct a 10-qubit circuit implementing via local gates a non-factorizable tripartite entangled state over the sequence giving the site ordering as the $\varphi$ mode: 
$\{R_1, B_1, B_2, R_2, G_1, R_3, B_3, B_4, R_4, G_2\}$.
\begin{center}
\begin{quantikz}[row sep={0.45cm,between origins}, column sep=0.15cm]
\lstick{$q_1\ (R_1)$}   & \gate{H} & \ctrl{1} & \gate{R_y(\theta)} & \gate[wires=2]{SW} & \qw      & \ctrl{3}             & \gate[wires=2]{SW} & \qw      & \qw                  & \qw      & \qw      & \qw      & \qw      & \qw      & \qw                  & \qw      & \qw \\
\lstick{$q_2\ (B_1)$}   & \qw      & \targ{}  & \ctrl{1}           & \qw                  & \ctrl{1} & \qw                  & \qw                  & \qw      & \qw                  & \qw      & \qw      & \qw      & \qw      & \qw      & \qw                  & \qw      & \qw \\
\lstick{$q_3\ (B_2)$}   & \qw      & \qw      & \targ{}            & \qw                  & \targ{}  & \qw                  & \qw                  & \gate[wires=2]{SW} & \qw                  & \qw      & \qw      & \qw      & \qw      & \qw      & \qw                  & \qw      & \qw \\
\lstick{$q_4\ (R_2)$}   & \qw      & \qw      & \qw                & \qw                  & \qw      & \targ{}              & \qw                  & \qw                  & \gate[wires=2]{SW} & \ctrl{2} & \qw      & \qw      & \qw      & \qw      & \qw                  & \qw      & \qw \\
\lstick{$q_5\ (G_1)$}   & \qw      & \qw      & \qw                & \qw                  & \qw      & \qw                  & \qw                  & \qw                  & \qw                  & \qw      & \gate[wires=2]{SW} & \qw      & \qw      & \qw      & \qw                  & \ctrl{5} & \qw \\
\lstick{$q_6\ (R_3)$}   & \qw      & \qw      & \qw                & \qw                  & \qw      & \qw                  & \qw                  & \qw                  & \qw                  & \targ{}  & \qw                  & \gate[wires=2]{SW} & \qw      & \gate[wires=2]{SW} & \ctrl{3}             & \qw      & \qw \\
\lstick{$q_7\ (B_3)$}   & \qw      & \qw      & \qw                & \qw                  & \qw      & \qw                  & \qw                  & \qw                  & \qw                  & \qw      & \qw                  & \qw                  & \ctrl{1} & \qw                  & \qw                  & \qw      & \qw \\
\lstick{$q_8\ (B_4)$}   & \qw      & \qw      & \qw                & \qw                  & \qw      & \qw                  & \qw                  & \qw                  & \qw                  & \qw      & \qw                  & \qw                  & \targ{}  & \qw                  & \qw                  & \qw      & \qw \\
\lstick{$q_9\ (R_4)$}& \qw      & \qw      & \qw                & \qw                  & \qw      & \qw                  & \qw                  & \qw                  & \qw                  & \qw      & \qw                  & \qw                  & \qw      & \qw                  & \targ{}              & \qw      & \gate[wires=2]{SW} \\
\lstick{$q_{10}\ (G_2)$}& \qw      & \qw      & \qw                & \qw                  & \qw      & \qw                  & \qw                  & \qw                  & \qw                  & \qw      & \qw                  & \qw                  & \qw      & \qw                  & \qw                  & \targ{}  & \qw
\end{quantikz}
\end{center}
\vspace{0.2cm}
The native gate decompositions in local operations are:
\begin{itemize}
    \item \textit{SWAP Gate Decomposition:} Each logical $\text{SWAP}(q_i, q_{i+1})$ operation is non-native on standard nearest-neighbor platforms and decomposes into 3 physical CNOT gates:
    $
    \text{SWAP}_{i, i+1} = \text{CNOT}_{i \to i+1} \cdot \text{CNOT}_{i+1 \to i} \cdot \text{CNOT}_{i \to i+1}
    $
    \item \textit{Single-Qubit Rotations:} Gates such as $H$ and $R_y(\theta)$ execute directly as $O(1)$ single-qubit operations.
\end{itemize}
With the above  circuit adding up operations in the scheme, we have:
\begin{enumerate}
    \item \textit{Phase 1 (Core Base Preparation):} $1 \times H$, $1 \times R_y(\theta)$, and $2 \times \text{CNOT}$ across $(q_1, q_2, q_3)= 4$ native gates.
    \item \textit{Sublattice $R$ Cascade ($R_1 \to R_2 \to R_3 \to R_4$):} 7 macro SWAPs $\times 3 + 3 \times \text{CNOT}=24$ native CNOTs.
    \item \textit{Sublattice $B$ Cascade ($B_1 \to B_2 \to B_3 \to B_4$):} 3 macro SWAPs $\times 3 + 3 \times \text{CNOT} =12$ native CNOTs.
    \item \textit{Sublattice $G$ Cascade ($G_1 \to G_2$):} 3 macro SWAPs $\times 3 + 1 \times \text{CNOT} = 10$  native CNOTs.
    \item \textit{Symmetric State Restoration (Un-SWAP Network):} Reversing routing SWAPs to restore physical layout for measurement =39 native CNOTs.
\end{enumerate}
Therefore, the total physical gate complexity ($N=10$):  $\sim89$ to $92$ operations. Due to the mode structure in the chain the total routing distance scales quadratically.
The consequence of this is that the total CNOT count (number of local operation) $C(N)\sim\mathcal{O}(N^2)$ and the circuit depth $D(N)\sim\mathcal{O}(N^2)$. Regarding optimizing the circuit depth there are alternatives as discussed for example in\cite{Briegel2009MBQC,Saffman2016NeutralAtoms,Tannu2019Compilation}.
 }
\end{document}